\documentclass[12pt,eqsecnum,nofootinbib,floats,preprint,aps,prd,floatfix,titlepage]{revtex4} 
\usepackage{epsfig}
\usepackage{amsmath}
\usepackage{amssymb}
\usepackage{graphics}
\usepackage{bm}

\begin{document}
\title{Negative modes of Coleman-De Luccia bounces}

\author{Hakjoon Lee}
\email{hl2406@columbia.edu}
\author{Erick J. Weinberg}
\email{ejw@phys.columbia.edu}
\affiliation{Physics Department, Columbia University, New York, New York 10027
\vskip 0.5in}

\begin{abstract}

We investigate the negative modes about Coleman-De Luccia bounces
governing vacuum transitions in de Sitter space, with the goal of
gaining physical insight into the various anomalous results associated
with these that have been reported in the literature.  For the case of
bounces with radii much less that the horizon distance $H^{-1}$ we
find two distinct regimes, distinguished by the magnitude of the
bubble nucleation rate $\Gamma$.  If $\Gamma/H^4 \gg 1$, then the
behavior of the modes contributing to the determinant factors in
$\Gamma$ is much as it is in flat spacetime, and the calculation of $\Gamma$
goes over smoothly to the flat spacetime calculation as the
gravitational coupling is taken to zero.  This is not the case if
$\Gamma/H^4 \ll 1$.  These two regimes correspond to the two possible
outcomes of de Sitter vacuum decay --- either a rapidly completed
transition or non-percolation and eternal inflation.  For bounces with
radii comparable to the horizon length, we confirm previous results
concerning anomalous negative modes with support on the bounce wall.
We also find further evidence supporting previous claims, based on
thin-wall arguments, of the absence of expected negative modes for a
class of bounces that arises when the initial and final vacua are nearly
degenerate.

\end{abstract}
\maketitle

\section{Introduction}
\label{introSec}

The process in which a false vacuum of a scalar field theory decays by
the nucleation of true vacuum bubbles was first studied in detail some 
time ago~\cite{Kobzarev:1974cp,Coleman:1977py}.  The bubble nucleation
rate has the exponential suppression typical of tunneling processes,
with the exponent being twice the WKB barrier penetration integral or,
equivalently, the Euclidean action of the bounce solution to the
Euclidean field equations~\cite{Coleman:1977py}.  By using path
integral methods to calculate the energy density of the false vacuum,
the pre-exponential factor can also be obtained~\cite{Callan:1977pt}.
This turns out to contain a factor of $[\det S_E''(\phi_{\rm
    bounce})]^{-1/2}$, where $S_E''$ denotes the second functional
variation of the Euclidean action.  An essential ingredient of this
calculation is the fact that $S_E''(\phi_{\rm bounce})$ has a single
negative eigenvalue, which arises from a mode corresponding to
expansion or contraction of the bounce.  Because of the square root,
this negative mode leads to a factor of $i$ that makes the false
vacuum energy complex, with its imaginary part interpreted in terms of
a rate of decay by bubble nucleation.  It is crucial here that there
is only a single negative eigenvalue and thus one factor of $i$; if,
for example, there were an even number of such eigenvalues, the
contribution to the false vacuum energy would be purely real.  Indeed,
it has been shown that the bounce with lowest action has one, and only
one, negative mode~\cite{Coleman:1987rm}.

Coleman and De Luccia (CDL)~\cite{Coleman:1980aw} proposed that this
formalism could be extended to include gravitational effects by adding
a Euclidean Einstein-Hilbert term to the action.  Although they did
not address the issue of the prefactor, it seems natural to assume
that it should be analogous to that for the non-gravitational case,
and that the dominant bounce should again have a single negative mode.
The matter is complicated by the fact that with gravity included the
fluctuations about the bounce enjoy a gauge freedom, corresponding
to the invariance under coordinate transformations of the underlying
theory.  A consequence of this gauge freedom is the existence of
constraints that must be imposed on the possible fluctuations.

A number of authors have studied the negative mode issue in the
context of bounces corresponding to bubble nucleation in a de Sitter
spacetime~\cite{Lavrelashvili:1985vn,Tanaka:1992zw,Garriga:1993fh,Tanaka:1999pj,Gratton:1999ya,
  Khvedelidze:2000cp,Lavrelashvili:1999sr,Gratton:2000fj,Dunne:2006bt}.
Some curious, troubling, and sometimes contradictory, results, often
depending on the choice of gauge, have been obtained.  Among these
are:

1) An action for fluctuations about a bounce with large radius that is
unbounded from below, suggesting an infinite family of negative modes
with support near the bubble wall.

2) Indications of a similar phenomenon for all de Sitter CDL bounces,
but with the additional negative modes having support on a region defined
by the horizon.

3) Arguments for the existence of a single negative mode, but subject to 
restrictions on the form of the bounce solution. 

4) Claims that there are no negative modes, and that the factor of $i$
comes from a proper rotation of contours in the Hamiltonian path
integral.

5) The absence of a negative mode in the thin-wall approximation for 
a certain class of bounces.
 
In our view, these works only address one part of the problem, namely
how to develop an algorithm that, at least in principle\footnote{Of
  course, even if the issues associated with negative modes are
  resolved, a complete calculation of the functional determinant would
  require an understanding of how to renormalize gravity.}, gives an
unambiguous result for the bubble nucleation rate.  There is another
aspect that should be addressed.  When anomalous or apparently
meaningless mathematical results are encountered, it is often the case
that they are indicative of a particular physical aspect of the
problem being studied.  Thus, one should look for a physical
understanding as to why these negative modes arise (or do not) in
particular cases, even if they can ultimately be eliminated from the
calculation.  Also, one should expect to find a smooth transition from
the curved spacetime calculation to the non-gravitational one in an
appropriate limit.  Ideally, this should include not just a continuous
evolution of the bounce and its action, but also a gradual
disappearance of the potential pathologies as one approaches the limit
in which $\kappa \equiv 8 \pi G =0$.  As we will see, this limit is
rather subtle.

Our goal in this paper is to make some progress toward this end by exploring 
the regions where the quadratic fluctuation action becomes negative.  We
work in a purely Lagrangian framework.  Rather than fixing a gauge, we
impose the requirement that the fluctuations obey the constraint, and
then write the quadratic fluctuation Lagrangian in terms of manifestly
gauge-invariant combinations of fields.  For fluctuations about a
bounce with O(4)-symmetry, which we assume, the objects of primary
interest are spherically symmetric scalar fluctuations, which can be
described by a single field $\chi$.  We examine the situations in
which $\chi$ can have negative modes, and study the form of those
modes as $\kappa$ is varied.

The remainder of this paper is organized as follows.  In
Sec.~\ref{tunnelSec} we review the formalism for treating vacuum decay
in both flat spacetime (i.e., with gravity ignored) and with
gravitational effects included.  In Sec.~\ref{negativeSec} we discuss
negative modes in flat spacetime.  In the course of this discussion we
describe an example where bounces with many negative modes turn out to
not only be physically meaningful, but to actually be dominant.
Section~\ref{perturb-sec} describes the perturbative expansion of the
scalar field and metric about an O(4)-symmetric CDL bounce and gives
the expression for the quadratic part of the fluctuation Lagrangian.
Next, in Sec.~\ref{negmode-sec}, we explore the various types of
negative modes that can arise from this Lagrangian.  Many of the
results that we present here are based on numerical explorations of
particular scalar field models.  In Sec.~\ref{multibounceSec} we
discuss the negative mode problem for multibounce solutions.  This is
almost a trivial aside in flat spacetime, but becomes more complex
with gravity included.  In fact, we will see that it is key to
understanding the limit in which gravitational effects disappear.
Finally, in Sec.~\ref{conclusionSec} we summarize our results and
discuss how they reflect the various modes of vacuum decay in de
Sitter spacetime.  There are two appendices containing some technical
details.

\section{Review of tunneling}
\label{tunnelSec}

\subsection{Tunneling in flat spacetime}
\label{flatTunnelSec}

The WKB approximation leads to an expression of the form
\begin{equation}
      \gamma = C\, e^{-B} 
\end{equation}
for the decay rate of a state that decays by tunneling through a potential energy
barrier.  For the case of a single particle in one dimension with a standard
kinetic energy,
\begin{equation}
      B = 2 \int_{x_1}^{x_2} dx \, \sqrt{2m[V(x) -E]}  \, ,
\label{tunnelInt}
\end{equation}
where $x_1$ and $x_2$ are the classical turning points that bound the particle's
path through the barrier.

With multiple degrees of freedom, the WKB approximation requires that
we consider all paths through the barrier and pick out the one along
which the one-dimensional tunneling exponent is a
minimum~\cite{Banks:1973ps,Banks:1974ij}.  This minimization problem
is equivalent to the problem of finding a stationary point of the
Euclidean action, which in turn is equivalent to finding a solution of
the Euclidean equations of motion
\cite{Coleman:1977py}.  The solution relevant for
tunneling, known as the bounce, starts at an initial configuration at
Euclidean time $\tau_{\rm in}$, runs through the barrier to an exit
point at some $\tau_0$, and then returns, in a ``$\tau$-reversed''
fashion, to $\tau_{\rm fin}$.  The intermediate Euclidean times give a
parameterization of the tunneling path through configuration space. If
the initial configuration is a local minimum of the potential energy
(i.e., a false vacuum), then $\tau_{\rm in}=-\infty$ and $\tau_{\rm
  fin}=\infty$.

For a scalar field theory with a standard Lagrangian the tunneling
exponent $B$ is equal to the difference between the Euclidean actions
of the bounce solution $\phi(x)$ and the uniform false vacuum.
Even if these actions are divergent, their difference,
\begin{eqnarray}
   B &=& S_E(\phi)  - S_E(\phi_{\rm fv})  \cr &=&
     \int_{-\infty}^\infty d\tau \int d^3x\, \left[
          \frac12 \left({d\phi \over d\tau}\right)^2 + \frac12 ({\bm \nabla}\phi)^2
        + U(\phi) - U(\phi_{\rm fv}) \right]  \, ,
\end{eqnarray}
is finite and well-defined.  Note that the tunneling integral and the
Euclidean action differ for paths that are not solutions of the
equations of motion.  In fact, although the bounce is a local minimum
of the integral through the potential energy barrier, it only
corresponds to a saddle point of $S_E$.

\begin{figure}[t]
\centering
\includegraphics[height=2.0in]{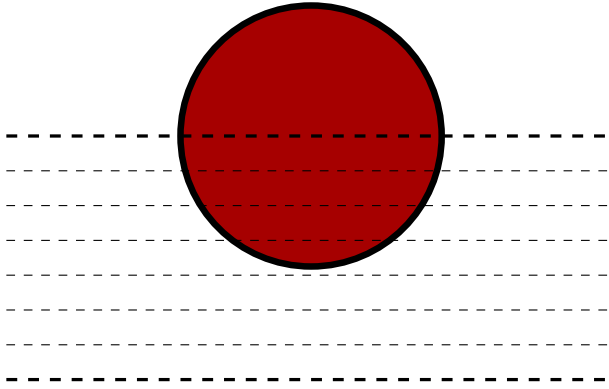}
\caption{Schematic view of a bounce in flat spacetime.  The shaded
  region represents the region of approximate true vacuum.  This is
  separated from the false vacuum exterior by the wall region
  represented by the solid black line.  (In actuality the field has an
  exponential tail and only reaches its exact false vacuum value at
  $\tau=-\infty$.)  The horizontal dashed lines represent
  hypersurfaces of constant $\tau$ that trace out a path in
  configuration space.  The heavy dashed lines correspond to the
  initial pure vacuum configuration and to the final field
  configuration at bubble nucleation.}
\label{flat-bounce}
\end{figure}

In a scalar field theory, with thermal and gravitational effects
ignored, the bounce describing tunneling from a false vacuum is a
solution on Euclidean $R^4$ that approaches the false vacuum value
$\phi_{\rm fv}$ as either $|{\bf x}|$ or $\tau$ tends to infinity.
With only minimal restrictions on the potential $U(\phi)$, it can be
shown that the bounce with the smallest action is
O(4)-symmetric~\cite{Coleman:1977th}.  Figure~\ref{flat-bounce}
illustrates such a bounce.

The prefactor $C$ can be obtained by a path integral
argument~\cite{Callan:1977pt}.  Consider the matrix element
\begin{eqnarray}
   \langle \phi_{\rm fv}| e^{-HT} | \phi_{\rm fv}\rangle &=&
      \int[d\phi({\bf x},\tau)]\, e^{-S_E[\phi]}  \cr
   &=& \sum_n e^{-E_nT} \langle \phi_{\rm fv}| n\rangle \langle n|
       \phi_{\rm fv} \rangle   \, .
\end{eqnarray}
Here the path integral is restricted to paths that begin and end on
the false vacuum configuration.  In the limit $T\to \infty$, the sum
over energy eigenstates in the second expression is dominated by the
state with the lowest energy among those that contribute to the matrix
element.  Identifying this state as the false vacuum, we have
\begin{equation}
  E_{\rm fv} =  -\lim_{T\to \infty} {1\over T} \ln \,
   \langle \phi_{\rm fv}| e^{-HT} | \phi_{\rm fv}\rangle  \, .
\label{EfvFromLog}
\end{equation}

The path integral is evaluated by expanding about its stationary points.
The first of these is the constant solution with $\phi(x) = \phi_{\rm fv}$ 
everywhere.  A Gaussian integral about this gives
\begin{equation}
  I_0 = [\det S_E''(\phi_{\rm fv})]^{-1/2}  \, e^{-S_E(\phi_{\rm fv})} \, ,
\end{equation}
where $S''_E$ denotes the second functional derivative of the Euclidean
action.  It is convenient to consider space to have a finite volume
$\Omega$ which is to be taken to infinity at the end of
the calculation, so that $S_E(\phi_{\rm fv}) = \Omega T \,U(\phi_{\rm fv})$.

Next, we have the bounce solution.  This would give a similar Gaussian
integral were it not for two factors.  First, the spectrum of
$S''_E(\phi_{\rm bounce})$ includes four zero modes, corresponding to
the freedom to translate the bounce in Euclidean space and time.  These
are handled by introducing collective coordinates specifying the
location of the center of the bounce.  Second, the spectrum also
contains a mode with negative eigenvalue, corresponding roughly (and
exactly, in the thin-wall limit) to expansion or contraction of the
bounce.  This negative mode leads to a factor of $i$ when the square
root of the determinant is taken.  The contribution to the path
integral can be written as
\begin{eqnarray}
     I_1 &=& \frac{i}{2}\,\Omega T \left|{\det' S_E''(\phi_{\rm bounce})] \over
          \det S_E''(\phi_{\rm fv})}\right|^{-1/2}  \,  J \,
       e^{-[S_E(\phi_{\rm bounce}) - S_E(\phi_{\rm fv})]}\, I_0  \cr
     &\equiv & i \Omega T K e^{-B} \,  I_0   \, .
\label{I1formula}
\end{eqnarray}
Here the factor of $\Omega T$ arises from integrating over the four
collective coordinates, $J$ contains the Jacobian factors associated with
the introduction of the collective coordinates, and the factor of 1/2
comes from a careful treatment of the negative mode.  Finally, the
prime on the bounce determinant indicates that the product of
eigenvalues is to be taken only over the nonzero eigenvalues.

There are also approximate stationary points corresponding to many 
well-separated bounces.  The $n$-bounce solution has an action 
\begin{equation}
      S_E(\phi_{n\rm -bounce}) = S_E(\phi_{\rm fv})
          + n \left[S_E(\phi_{\rm bounce}) - S_E(\phi_{\rm fv}) \right]
        = S_E(\phi_{\rm fv})  + n B  \, .
\end{equation} 
The integration over the collective coordinates gives a factor of
$(\Omega T)^n/n!$, with the $n!$ entering because the
bounces are indistinguishable.  Similarly, the factor of $J$ becomes
$J^n$.  Only the determinant factors remain to be considered.

Outside the wall region the bounce solution 
rapidly approaches the pure false vacuum, with $|\phi_{\rm bounce}(x)
- \phi_{\rm fv}|$ decreasing exponentially with distance.  Thus, we
can imagine evaluating the determinant for the one-bounce case by
dividing Euclidean space into a large (compared to the bounce radius)
region enclosing the bounce, and the remainder.  The full determinant
is then the product of the contributions of the two regions.  In the
latter region the bounce is exponentially close to the false vacuum
and the contributions to the bounce and the false vacuum determinants
are essentially equal.  In the region containing the bounce the
determinants corresponding to $\phi_{\rm bounce}$ and $\phi_{\rm fv}$
differ precisely by the ratio that appears in Eq.~(\ref{I1formula}).

For an $n$-bounce configuration, with the bounces all well separated,
the Euclidean spacetime can be divided into $n$ regions, each containing 
one bounce, and the remainder, in which the field is essentially equal 
to its false vacuum region.  The full determinant is the product of the
contributions from each of these regions.  The net result is that the 
contribution to the path integral from $n$-bounce configurations is 
\begin{equation} 
     I_n = {1\over n!} \left( i \Omega T K e^{-B} \right)^n I_0 \, .
\end{equation}
Summing the contributions from all values of $n$ gives
\begin{eqnarray}
     I &=& I_0 \sum_{n=0}^\infty {1\over n!} 
          \left( i \Omega T K e^{-B} \right)^n  \cr
      &=& I_0 \exp\left[ i \Omega T K e^{-B}  \right] \, .
\label{Isum}
\end{eqnarray}
Taking the logarithm and using
Eq.~(\ref{EfvFromLog}) gives
\begin{equation}
      E_{\rm fv} =  -\lim_{T\to \infty} \left( {\ln \,I_0\over T}\right)
               - i\Omega K e^{-B} \, .
\end{equation}
The first term on the right-hand side is real, but the second 
is imaginary, making the false vacuum energy complex.  As usual,
we interpret this as a signal that the false vacuum is unstable, 
with a decay rate   
\begin{equation}
     \gamma = - 2 \,{\rm Im}\, E_{\rm fv} = 2 \Omega K e^{-B} \, .
\end{equation}
This is proportional to $\Omega$, corresponding to the fact that 
a bubble can nucleate anywhere in space.  The bubble nucleation rate
per unit volume is 
\begin{equation}
     \Gamma = 2 K \, e^{-B}  \, .
\end{equation}

This calculation implicitly assumes that in the multibounce solutions
the individual bounces are separated by distances large relative to
the bounce four-volume ${\cal V}_b$.  To test this dilute-gas
approximation, we note that the sum in Eq.~(\ref{Isum}) is dominated by the
terms with
\begin{equation}
     n \approx \Omega T K e^{-B}  \, .
\label{dominantN}
\end{equation}
For the approximation to be valid, we must require that the volume
occupied by these $n$ bounces be much less than the total Euclidean spacetime
volume $\Omega T$; i.e.,
\begin{equation}
     {\cal V}_4 K e^{-B}  \ll 1  \, ,
\end{equation}
where ${\cal V}_4$ is the four-dimensional volume of the bounce.  Dimensional 
arguments suggest that ${\cal V}_4 K$ is typically of order unity, in which 
case this condition reduces to a lower bound on $B$.

A useful illustrative example is given by the thin-wall approximation,
which applies in the limit where the difference between the false and
true vacuum values of the potential, $\epsilon \equiv U(\phi_{\rm fv})
- U(\phi_{\rm tv})$, is sufficiently small relative to the surface
tension $\sigma$ of the bubble~\cite{Coleman:1977py}.  The bounce can
then be obtained by considering O(4)-symmetric configurations in which
a true vacuum region of radius $R$ is separated from the exterior
false vacuum by a thin wall with action per unit area $\sigma$.  The
total Euclidean action, less that of the pure false vacuum, is
\begin{equation}
    S_E(R) = 2\pi^2 \sigma R^3 - {\pi^2 \over 2}\, \epsilon R^4 \, .
\end{equation}
The stationary point of this action at 
$\bar R \equiv 3 \sigma/\epsilon$ gives the bounce radius. 
Both the bounce radius and 
the tunneling exponent
\begin{equation}
    B = {27 \pi^2\over 2}\, {\sigma^4 \over \epsilon^3} 
\end{equation}
tend to 
infinity in the $\epsilon \to 0$ limit of degenerate vacua.  The 
fact that 
\begin{equation}
   \left. {d^2 S_E \over  d R^2}\right|_{\bar R}
        = -{18 \pi^2 \sigma^2 \over \epsilon}
\label{TWAsecondDeriv}
\end{equation}
is negative shows that the bounce is a maximum of the action among the
one-parameter family of thin-wall configurations, and thus a saddle
point of the action, with at least one negative eigenvalue, on the full 
configuration space.

Finally, recall that 
the bounce is supposed to describe a family of configurations interpolating
between the initial pure vacuum state, at $\tau=-\infty$, and a state of
equal potential energy containing a bubble, on a constant $\tau$ slice through
the center of the bounce.   In the thin-wall limit the difference in the 
static potential energy between these two configurations is 
\begin{equation}
     \Delta E = 4\pi R^2 \sigma - \frac43 \pi R^3 \epsilon  \, .
\end{equation}
This does indeed vanish if $R = \bar R$.

The above discussion readily generalizes to the case of nonzero
temperature $T$~\cite{Linde:1981zj}.  In this case $\Gamma$ is
obtained from the imaginary part of the free energy, rather than the
energy, of the metastable false vacuum.  The Euclidean path integral
is then over paths that are periodic in Euclidean time with period
$\beta=1/T$.  Furthermore, the classical Euclidean equations leading
to the bounce must be obtained using the finite-temperature effective
potential.  Just as at $T=0$, the path integral includes contributions
not just from a single bounce, but also from all multibounce
solutions.  Summing over these exponentiates the single bounce
contribution, and the nucleation rate can be read off from the
exponent.

\subsection{Including gravitational effects}

Coleman and De Luccia~\cite{Coleman:1980aw} argued that gravitational
effects could be incorporated by adding a Euclidean Einstein-Hilbert
term to the action, so that
\begin{equation}
     S_E = \int d^4x \sqrt{g} \left[-{1 \over 2\kappa}R
       +\frac12\,g^{ab} \partial_a\phi\,\partial_b \phi +U(\phi) \right]
         + S_{\rm bdy}  \, ,
\end{equation}
where $\kappa = 8\pi G$ and $S_{\rm bdy}$ is the Euclidean version of
the Gibbons-Hawking boundary term~\cite{Gibbons:1976ue}.  

Although it has not been proven that the bounce of minimum
action continues to be O(4)-symmetric 
when gravitational effects are included, this
is widely believed to be true.  If one assumes this to be the case,
the Euclidean metric can be written as
\begin{equation}
   ds^2 = N(\xi) d\xi^2 + \rho(\xi)^2 d\Omega_3^2  \, ,
\end{equation}
where $d\Omega_3^2$ is the usual metric on the unit three-sphere.  It
is convenient to choose the origin of $\xi$ to be a zero of $\rho$, so
that $\rho(0)=0$.  The curvature scalar is 
\begin{equation}
     R = {6 \over N \rho^2}\left(N-{\dot\rho}^2 - \rho\ddot\rho\right)
        + {3 \dot\rho\dot N\over \rho N^2} \, ,
\end{equation}
with overdots denoting derivatives with respect to $\xi$.

With the scalar field $\phi$ depending only on 
$\xi$, the Euclidean action takes the form 
\begin{equation}
    S_E = 2\pi^2 \int d\xi \sqrt{N} \left\{ \rho^3\
    \left[{1\over 2N}\,{\dot\phi}^2  + U(\phi)\right]
   +\frac3\kappa \left[{1\over N}\left(\rho^2\ddot\rho + \rho{\dot\rho}^2\right)
           -\rho  -{\rho^2\dot\rho \dot N \over 2 N^2}  \right] \right\}
   + S_{\rm bdy} \, .
\end{equation}
An integration by parts to remove the second derivative term recasts
this as
\begin{equation}
   S_E = 2\pi^2 \int d\xi\sqrt{N} \left\{ \rho^3\left[{1 \over 2N}\,{\dot\phi}^2
          + U(\phi)\right]
    -\frac3\kappa \left( {\rho{\dot\rho}^2 \over N} +\rho\right) \right\} \, ,
\end{equation} 
with the boundary terms from the integration by parts precisely canceling
$S_{\rm bdy}$.

We note that the action does not contain any derivatives of $N$.
Consequently, variation with respect to $N$ yields a constraint
equation (which is in fact the $G_{\xi\xi}$ Einstein equation),
\begin{equation}
     0= \rho^3\left({{\dot\phi}^2\over 2N} - U \right) 
  -{3\over \kappa}\left({\rho{\dot\rho}^2 \over N} -\rho \right)  \, .
\label{rhoN-eq}
\end{equation}
The existence of this constraint is related to the freedom to make a 
coordinate transformation to redefine $\xi$.  We make use of this gauge
freedom to set $N(\xi)=1$.  Having done so, we can write the constraint
equation as
\begin{equation} 
      {\dot\rho}^2 = 1 + {\kappa \over 3} \, \rho^2 
           \left(\frac12 {\dot\phi}^2 -  U\right)  \, .
\label{rho-eq}
\end{equation}
Varying the action with respect to $\phi$ gives the scalar
field equation of motion
\begin{equation}
 \ddot\phi + {3 \dot\rho\over \rho}\dot\phi = {dU \over d\phi} \, .
\label{phi-eq}
\end{equation}
Equations~(\ref{rho-eq}) and (\ref{phi-eq}) are a complete set of
field equations.  Variation of the action with respect to $\rho$ does
not yield an independent equation, reflecting the fact that in
spherically symmetric configurations the 
gravitational dynamics is determined completely by the matter
distribution.  However, differentiating
Eq.~(\ref{rho-eq}) and then using Eq.~(\ref{phi-eq}) yields the 
useful identity
\begin{equation}
  \ddot \rho = - \frac{\kappa}{3} \, \rho\left(\dot\phi^2 + U \right) \, .
\end{equation}

For bubble nucleation in a Minkowski or anti-de Sitter false vacuum
the bounce solution has the topology of $R^4$, just as in the
non-gravitational case.  De Sitter spacetime, the case on which we will
focus, is different.  In this
case $\rho$ always has a second zero, at a value $\xi=\xi_{\rm max}$.
The bounce is then topologically a four-sphere, with a finite
four-volume of order $H^{-4}$, where
\begin{equation}
     H = \sqrt{{\kappa \over 3}\, U(\phi_{\rm fv}) }
\label{HubbleDef}
\end{equation}
is the Hubble parameter of the false vacuum.\footnote{To simplify
  notation we will omit the subscript on $H$ when it refers to the
  false vacuum; when we need to refer to the true vacuum Hubble
  parameter this will be indicated by a subscript.}  In particular, for
the solution of Eqs.~(\ref{rho-eq}) and (\ref{phi-eq}) describing the
pure false vacuum, with $\phi= \phi_{\rm fv}$ everywhere, the metric
is the standard round metric on the four-sphere, with
\begin{equation}
     \rho = H^{-1} \sin(H \xi)  \, .
\label{SphereRho}
\end{equation}
Its Euclidean action, which must be subtracted from that of the bounce 
to obtain $B$, is
\begin{equation}
    S_E =  - {24\pi^2 \over \kappa^2 \, U(\phi_{\rm fv})}
       = - {8\pi^2 \over \kappa} \, H^{-2} \, .
\end{equation}

In our discussions it will be useful to distinguish two classes
of bounce solutions:

a) Small bubble bounces:  The difference between the values of
the potential at the false vacuum and at the top of the barrier defines
a mass scale $\mu$ via
\begin{equation}
     U_{\rm top} - U_{\rm fv} \equiv \mu^4  \, .
\label{mudef}
\end{equation}
If $\mu \ll M_P$ and the difference between the true and false vacuum values
of $U$ is not too small, then the bounce describes the nucleation of bubbles
whose radius $\bar \rho$ at nucleation is much smaller than $H^{-1}$.  
If $\xi \ll H^{-1}$, then $\rho(\xi) \approx \xi$ and $\phi(\xi)$ differs
only slightly from the corresponding flat-space bounce. For larger $\xi$, on 
the other hand, 
\begin{equation}
    \rho =  H^{-1} \sin[H (\xi + \Delta)]  \, ,
\label{rhoInFV}
\end{equation} 
where $\Delta = O(\mu/M_P^2)$. The scalar field in this large-$\xi$ region is
exponentially close to its false vacuum value, with $|\phi - \phi_{\rm fv}|
\sim \mu e^{-\mu(\xi - \bar\xi)}$.

b) Large bubble bounces: Bubbles of size comparable to $H^{-1}$ can
arise, even with $\mu \ll M_P$, if $\epsilon$ is sufficiently small.
The flat-space thin-wall results are replaced by~\cite{Parke:1982pm}
\begin{equation} 
     {1 \over  \bar\rho^2}  = H^2 + \left( {\epsilon \over 3 \sigma}
       - {\kappa \sigma \over 4} \right)^2  \, .
\label{TWArhoBar}
\end{equation}
As $\epsilon$ is decreased, the bubble radius increases until it
reaches a maximum value, $\bar\rho = H^{-1}$, when the quantity in
parentheses vanishes.  Until this point, the false vacuum region is
larger than a hemisphere, while the true vacuum occupies less than a
hemisphere.  With a further decrease in $\epsilon$, $\bar\rho$ begins
to decrease, and the true and false vacuum regions are both less than
hemispheres.  In contrast to the flat-space thin-wall bounce, whose
radius and action tend to infinity in the $\epsilon \to 0$ limit, both
$\bar\rho$ and the bounce action remain finite in this limit.  To
distinguish between the two regimes, we will denote bounces as being
type A or type B, according to whether the false vacuum region is
greater than or less than a hemisphere.   

We saw in Eq.~(\ref{TWAsecondDeriv}) that with gravity ignored the
bounce was a maximum, as a function of $R$, of the
thin-wall-approximation action.  Similarly, for type A bounces the
$\bar\rho$ of Eq.~(\ref{TWArhoBar}) gives a maximum of the thin-wall
action, and therefore a saddle point of the full action
functional~\cite{Coleman-Steinhardt}.  However, for type B thin-wall
solutions the second derivative of the action with respect to
$\bar\rho$ is positive, corresponding to a minimum.

We can generalize the distinction between type A and type B
solutions beyond the thin-wall limit.  In the thin-wall limit the
maximum of $\rho$ in a type B bounce occurs at the bubble wall.  More
generally, we will say that a bounce is type B if the maximum of
$\rho$ occurs within the wall region, and type A otherwise.  (Because
the beginning and end of the wall are not precisely defined, there
will still be some ambiguity concerning borderline cases.)

\begin{figure}[t]
\centering
\includegraphics[height=2.0in]{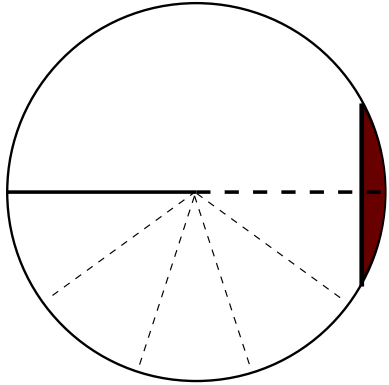}
\caption{Schematic view of a Coleman-De Luccia bounce, with the
  coordinate $\xi$ running horizontally, from $\xi=0$ on the right to
  $\xi_{\rm max}$ on the left.  As in Fig.~\ref{flat-bounce}, the
  shaded region represents the region of approximate true vacuum and
  the unshaded region that of approximate false vacuum.  A path
  through configuration space, starting from the solid horizontal line
  and ending on the heavy dashed line, is given by the lighter dashed
  lines.  Each of these lines represents a horizon volume that is
  topologically a three-ball with its center at the edge of the
  figure and its two-sphere horizon boundary at the center of the
  figure.}
\label{CDL-bounce}
\end{figure}

The fact that the bounce topology is compact raises questions: Why is
the volume finite? How can the bounce be understood as a sequence of
configurations forming a tunneling path through a potential energy
barrier?  It was argued in~\cite{Brown:2007sd} that the bounce should
be understood as mediating a transition in a de Sitter horizon volume
(giving a finite three-volume $\sim H^{-3}$) at a finite de Sitter
temperature $T_{\rm dS} = H/2\pi$ (giving periodicity $\sim H^{-1}$ in
a fourth Euclidean coordinate).  In this picture the path through
configuration space is given by a series of radial slices, each of
which represents a three-dimensional ball, as shown in
Fig.~\ref{CDL-bounce} (see also~\cite{Tanaka:1992zw}).  The boundaries
of the various three-balls, which are two-spheres at the horizon, all
meet and are identified.  The initial and final slices are the two
horizontal slices in this figure.  In the non-gravitational case, the
bounce connects initial and final configurations that are each turning
points, with vanishing derivatives with respect to the Euclidean time,
that can be continued to real time configurations with vanishing time
derivatives.  With gravitational effects included, the initial and
final slices each correspond to a configuration on a horizon volume
with instantaneously static metric and scalar field.  In
Appendix~\ref{horizonApp} we verify explicitly that these limiting
configurations are indeed bounded by horizons.

\section{Negative modes}
\label{negativeSec}

The occurrence of a negative eigenvalue in the spectrum of fluctuations
about the bounce was an essential ingredient in the path integral
derivation of the flat-space nucleation rate.  It gave rise to the
factor of $i$ in Eq.~(\ref{I1formula}), which led to the bounce terms
being a contribution to the imaginary part of the false vacuum energy.

The existence of the
negative mode is readily established in flat spacetime.  An O(4)-symmetric 
bounce obeys
\begin{equation}
      {d^2\phi \over ds^2} + {3 \over s}{d\phi \over ds}
           = U'(\phi)  \, ,
\label{flatScalarEq}
\end{equation}
where $s = \sqrt{\tau^2 + {\bf r}^2}$.  The eigenmodes about this bounce 
satisfy 
\begin{equation}
    \left[- \Box_E + U''(\phi_b) \right] \eta = \lambda \eta \, .
\label{BoxEeq}
\end{equation}
The spherical symmetry of the underlying bounce allows us to decompose
the modes as products of a radial function and a four-dimensional
spherical harmonic, and to recast Eq.~(\ref{BoxEeq}) into the form of a
Schroedinger equation, with the angular derivatives giving rise to an
angular momentum barrier term.  We know that the breaking of
translational invariance gives rise to four zero modes that transform as
the components of a four-dimensional vector.  Standard arguments then
show that there must be a rotationally invariant scalar mode with a 
lower, and therefore negative, eigenvalue $\lambda$ that obeys
\begin{equation}
  - {d^2\eta \over ds^2} - {3 \over s}{d\eta \over ds} 
         + U''(\phi_b) \eta = \lambda \eta  \, .
\label{flat-mode-eq}
\end{equation} 

This is readily verified for the case of a thin-wall bounce with
radius $\bar R$.  Differentiating Eq.~(\ref{flatScalarEq}) with
respect to $s$ gives
\begin{equation}
     \left[ -{d^2 \over ds^2}- {3 \over s}{d \over ds}  
       + U''(\phi_b) \right] {d\phi_b \over ds} 
     = -{3\over s^2} {d\phi_b \over ds}  \, .
\label{flat-neg-mode}
\end{equation}
In the thin-wall limit $d\phi/ds$ vanishes except in a narrow
region where $s \approx \bar R$, and so we see that there is an 
eigenfunction $\eta(s) \approx d\phi/ds$ with negative eigenvalue
$\lambda \approx -3/\bar R^2$.

What about multiple negative modes?  The determinant factor would
acquire a factor of $i$ for each such mode, and so would be real if
there were an even number of negative modes.  Three such modes
would give an imaginary determinant factor, but with the wrong 
sign.  Although five modes, or any larger number equal to $1 \mod 4$, would 
give an imaginary part of the right sign, this would seem to be an 
unusual constraint on the bounce.

The situation was clarified by Coleman~\cite{Coleman:1987rm}, who
showed that if a bounce has multiple negative modes then the
corresponding tunneling path through configuration space is a saddle
point, rather than a minimum, of the barrier penetration integral.  As
an explicit illustration of this, consider a theory with a single
scalar field whose potential has three minima, at $\phi_A < \phi_B <
\phi_C$, of increasing depth, $U_A > U_B > U_C$~\cite{Tye:2009rb}.  It
is convenient to choose the zero of energy so that $U_A =0$.  Let us
assume that the energy density differences
\begin{eqnarray} 
      \epsilon_1 &\equiv&  U_A - U_B  \, , \\
      \epsilon_2 &\equiv&  U_B - U_C  \, ,
\end{eqnarray} 
and the corresponding surface tensions $\sigma_1$ and $\sigma_2$ are
such that the thin-wall approximation is valid for both the A-to-B and
the B-to-C transitions.   The corresponding thin-wall bubble radii are
\begin{equation}
     \bar R_j = {3 \sigma_j \over \epsilon_j}\, , \qquad j=1,2 \, .
\end{equation}

Now consider a ``double-bounce'' configuration in which a spherical
C-vacuum region of radius $R_2$ is centered within a spherical
B-vacuum region of radius $R_1$, which itself is surrounded by an
A-vacuum background.  This is clearly a solution of the Euclidean
field equations if $R_1 =\bar R_1$ and $R_2 =\bar R_2$.  It
is also clear that there are two negative modes, corresponding to
independent variations of $R_1$ and $R_2$.  

With a single spherical thin-wall configuration $\bar R$ is the unique
radius consistent with having the four-dimensional fields interpolate
between the false vacuum, at $\tau=-\infty$, and a three-dimensional
configuration of equal energy containing a single bubble, at $\tau=0$.
With two nested three-spheres there is a one-parameter family of
configurations satisfying this equal energy condition, with their radii
constrained to obey
\begin{equation}
  0= 4\pi \sigma_1  R_1^2 -{4\pi \over 3} \epsilon_1
   R_1^3 
    + 4\pi \sigma_2 R_2^2 -{4\pi \over 3} \epsilon_2
  R_2^3 \, .
\end{equation}  
For each choice of $R_1$ and $R_2$ obeying this condition one can
calculate the potential energy\footnote{For some choices of parameters
  and of the $R_j$
  there is a range of $\tau$ in which the potential energy is less
  than that of the original false vacuum. The cases with such
  classically allowed regions were the focus of~\cite{Tye:2009rb}, but are
  not of relevance for the present discussion.}
\begin{equation}
     V(\tau) = \int d^3x \left[\frac12 ({\bm \nabla} \phi)^2 + U(\phi)\right]
\end{equation}
as a function of $\tau$ and then calculate the field theory
generalization of the barrier penetration integral of
Eq.~(\ref{tunnelInt}).  As expected, this has a minimum when $R_2=0$
and $R_1= \bar R_1$ (i.e. the ordinary A-to-B bounce).  On the other
hand, the choice $R_1 =\bar R_1$ and $R_2 =\bar R_2$ that gives the
double-bounce solution turns out to maximize this tunneling integral.  Thus, we
would not expect this double-bounce solution to correspond, even with
a reduced rate, to a new mode of bubble nucleation.

However, suppose that the parameters are such that the preferred radius, $\bar R_2$,
of the B-to-C bounce is much less than $\bar R_1$.  There will then be 
solutions in which a single A-to-B bounce contains many B-to-C bounces.
In evaluating the path integral, we would then have to sum not only over 
all numbers of A-to-B bounces, as in Eq.~(\ref{Isum}), but also over the
numbers of B-to-C bounces within each of these bubbles.   This would lead to 
a double sum formula
\begin{eqnarray}
   I &=& I_0 \sum_{n=0}^\infty  {1\over n!}\, \left[i\Omega TK_1e^{-B_1}
  \sum_{k=0} {1 \over k!}\left(i {\cal V}_4 K_2 e^{-B_2}\right)^k \right]^n \cr
   &=& I_0 \sum_{n=0}^\infty  {1\over n!}\, \left[i\Omega TK_1e^{-B_1}
  \sum_{k=0} {1 \over k!}\left({i\over 2}\, {\cal V}_4 \Gamma_2\right)^k \right]^n \, ,
\end{eqnarray}
where 
\begin{equation}
   {\cal V}_4  = {81 \pi^2 \over 2} \, \left(\sigma \over \epsilon \right)^4
\end{equation}
is the four-volume of the A-to-B bounce and
$\Gamma_2$ is the bubble nucleation rate for the B-to-C transition.
Because $\Omega T$ is understood to be eventually taken to infinity,
the upper limit on the sum over $n$ can be taken to be infinity, allowing 
us to convert the sum to an exponential.  By contrast, ${\cal V}_4$ is fixed
and finite, leading us to consider two regimes.  If ${\cal V}_4 \Gamma_2$
is much less than unity, the sum over $k$ is dominated by the $k=0$ term,
the A-to-B bounces containing B-to-C sub-bounces can be ignored, and there is 
no significant correction to $\Gamma_1$.  On the other hand, if 
${\cal V}_4 \Gamma_2$ is much greater than unity, the sum over $k$ can 
be approximated by an exponential, which has the effect of replacing $B_1$
by~\cite{Czech:2011aa} 
\begin{equation}
    \tilde B_1 =  B_1 - {i\over 2} {\cal V}_4 \Gamma_2 
    = {27 \pi^2 \over 2} \, {\sigma_1^4 \over \epsilon_1^3} 
       \left( 1 - {3i\over 2} \, {\Gamma_2 \over  \epsilon_1}\right) \, .
\end{equation}
To leading order, this is the same as making the replacement
$\epsilon_1 \rightarrow \epsilon_1 + i\Gamma_2/2$ or, in other words,
as if we had given the energy density of vacuum B an imaginary part
$\Gamma_2/2$.  This describes a situation in which vacuum B is
sufficiently unstable that bubbles of B already contain tiny bubbles
of C at the time that they nucleate. Thus, although though the
double-bounce with two negative modes has little physical relevance,
the solutions with many sub-bounces and hence many negative modes are
physically meaningful.\footnote{In a two-field
  model~\cite{Balasubramanian:2010kg} a similar mechanism can result
  in the wall tension acquiring a small imaginary
  part~\cite{Czech:2011aa}.}  Although they
  have a somewhat higher action than the unadorned single bounce, this
  is outweighed by the fact that they are more numerous.  In essence,
  it is a case of ``entropy'' overcoming ``energy''.

\section{Perturbative expansion about a CDL bounce}
\label{perturb-sec}

In this section we consider the expansion of the Euclidean action
about an O(4)-symmetric bounce solution of Eqs.~(\ref{rho-eq}) and
(\ref{phi-eq}), and obtain the contribution from the terms quadratic
in fluctuations about the bounce.  We require that these fluctuations
obey the constraints that follow from the freedom to make coordinate
transformations.  To exclude fluctuations that are purely coordinate
transformations, we write the quadratic action in terms of explicitly
gauge-invariant quantities.

It is convenient to exploit the O(4) symmetry by expanding the
fluctuations in terms of O(4) harmonics.  The normal modes can be
classified as scalar, vector, or tensor.  Because the only matter
source is a scalar field, it suffices for us to focus on the scalar
modes.  We begin by considering the most important case, the
O(4)-symmetric zero angular momentum modes.

\subsection{O(4)-symmetric fluctuations}

With O(4) symmetry retained, we can write the metric as
\begin{equation}
   ds^2= [1+2A(\xi)] d\xi^2+\rho(\xi)^2[1+2\Psi(\xi)]d\Omega_3 ^2 \, ,
\label{ZeroEllMetric}
\end{equation}
corresponding to the perturbations
\begin{eqnarray}
    N(\xi) &\rightarrow& 1 + A(\xi)  \, ,  \cr
    \rho(\xi) &\rightarrow& \rho(\xi)[1 + \Psi(\xi)] \, , 
\label{zeroEllPert}
\end{eqnarray}
and define the perturbed scalar field
\begin{equation}
    \phi(\xi)\rightarrow \phi(\xi)+\Phi(\xi) \, .
\label{PhiPert}
\end{equation}

Next we expand the total action about the background solution. The
first-order correction to the action vanishes by the background
equations of motion.  The quadratic terms are given by
\begin{equation}
  S^{(2)}_E=2\pi^2 \int  L^{(2)}_E(\Phi,\Psi,A;\dot{\Phi},\dot{\Psi}) d\xi \, ,
\end{equation}
where
\begin{eqnarray}
     L^{(2)}_E(\Phi,\Psi,A;\dot{\Phi},\dot{\Psi})=
    -\frac{3}{\kappa}\rho^3 \dot{\Psi}^2+
    \frac{3}{\kappa}\rho\Psi^2+\frac{1}{2}\rho^3\dot{\Phi}^2
    +\frac{1}{2}\rho^3 U''\Phi^2-3\rho^3 \dot{\phi}\dot{\Psi}\Phi \,\cr
       +\left(-\rho^3\dot{\phi}\dot{\Phi}+\rho^3 U' \Phi 
      +\frac{6}{\kappa}\dot{\rho}\rho^2\dot{\Psi}
    +\frac{6}{\kappa}\rho \Psi\right)A-\frac{3}{\kappa}\rho  Q A^2  \, .
\label{zeroEllL2}
\end{eqnarray}
Here we have defined\footnote{This definition agrees with that in
  Refs.~\cite{Lavrelashvili:1985vn} and \cite{Tanaka:1992zw}, but
  differs from the quantity denoted by $Q$ in
  Refs.~\cite{Khvedelidze:2000cp} and~\cite{Dunne:2006bt}. For 
$l=0$ modes the latter two references define $Q=1 - \kappa \rho^2 \dot\phi^2/6$.}
\begin{equation}
        Q= \left(1 - {\kappa \rho^2 U \over 3}\right)
         = \dot\rho^2 -\frac{\kappa \rho^2\dot{\phi}^2}{6} \, ,
\label{Qdef}
\end{equation}
with the two expressions being equal because of Eq.~(\ref{rho-eq}).

The theory is invariant under coordinate transformations.  For
perturbations about a background solution, the coordinate
transformation $\xi \rightarrow \xi + \alpha(\xi)$ gives the
infinitesimal gauge transformation
\begin{equation}
    \delta_G \Phi = \dot{\phi}\alpha,
    \quad \delta_G \Psi=\frac{\dot{\rho}}{\rho}\alpha,
    \quad \delta_G A = \dot{\alpha}  \, .
\label{O4gaugeTrans}
\end{equation}
This leaves $L_E^{(2)}$ unchanged, up to a total derivative (see
Appendix~\ref{mode-app}), and so is an invariance of the action.

Closely related to this gauge invariance is the existence of a
constraint that follows from the fact that $L^{(2)}_E$ does not
contain derivatives of $A$.  Requiring that $\delta L^{(2)}_E/\delta
A$ vanish gives
\begin{equation}
    0 = {\cal C}^{(1)} \equiv  \frac{\kappa \rho^2}{6}
\left(\dot{\phi}\dot{\Phi}-U'\Phi\right)
   -\left(\rho \dot{\rho}\dot{\Psi}+\Psi\right)+ Q A \, .
\end{equation}
This is nothing more than the linear term in the expansion of the
constraint arising from the variation $N(\xi)$, and could have been
obtained by substituting Eqs.~(\ref{zeroEllPert}) and (\ref{PhiPert})
into Eq.~(\ref{rhoN-eq}).  Using this constraint we can eliminate $A$
from Eq.~(\ref{zeroEllL2}).  After some tedious calculations,
described in Appendix~\ref{mode-app}, we obtain the remarkably simple
expression
\begin{equation}
       L^{(2)}_E(\chi;\dot{\chi})=\frac{\rho^3}{2Q}\dot{\chi}^2+
         \frac{\rho^3}{2Q}f(\rho,\phi)\chi^2  \, ,
\label{g-inv-L2}
\end{equation}
where we have defined the gauge-invariant quantity
\begin{equation}
      \chi \equiv \dot\rho \Phi-\rho\dot\phi\Psi 
\label{ZeroEllChi}
\end{equation}
and 
\begin{equation}
    f(\rho ,\phi) = U''+\frac{\kappa \rho^2 U'^2}{3 Q}
    +\frac{\kappa \rho\dot{\phi} U'}{3\dot{\rho} Q}
    +\frac{2\kappa \dot\phi^2}{3} - \frac{\kappa\rho\dot\phi U'}{\dot\rho}    
     -\frac{4\kappa U}{3}  -{\ddot\rho \dot Q\over  \dot\rho Q}  \, .
\label{fForL2}
\end{equation}
Our expression for the second-order Lagrangian
$L^{(2)}_E$ is now manifestly gauge-invariant.
One can check that the constraint
${\cal C}^{(1)}$ is likewise gauge-invariant.

\subsection{Non-spherically-symmetric perturbations}

The O(4)-symmetric perturbations of the metric are
completely described by the two functions $A$ and $\Psi$.  
Less symmetric perturbations require additional fields.
For arbitrary angular momentum the perturbed metric
can be written as
\begin{eqnarray}
     ds^2 &=& [1+2 A_l(\xi)Y_l(\Omega)]\, d\xi^2
   + B_l(\xi) \nabla_a Y_l(\Omega)\, d\xi \,dz^a \, \cr
  &&+\rho(\xi)^2\left\{\bar{g}_{ab} [1+2\Psi_l(\xi)Y_l(\Omega)]
      +2C_l(\xi)k^{-2}\left(\nabla_a \nabla_b +\frac{k^2}3\bar{g}_{ab}\right) 
       Y_l(\Omega)\right\}dz^a dz^b \, .  \cr &&
\label{generalPertMetric}
\end{eqnarray}
Here $\bar{g}_{ab}$ is the standard round metric on the three-sphere with 
coordinates $z^a$ and $\nabla_a$ is the corresponding covariant derivative.
The spherical harmonics 
$Y_l(\Omega)$ are eigenfunctions of the Laplacian on the three-sphere, with
\begin{equation}
   \Delta Y_l(\Omega)=-l(l+2)\, Y_l(\Omega)=-k^2 \,Y_l(\Omega) \, .
\end{equation}
For simplicity of notation we have defined $k^2=l(l+2)$ and have suppressed 
the indices corresponding to the $(l + 1)^2$ degeneracy of the modes and spherical
harmonics of angular momentum $l$.  In the following we will also omit the subscript
$l$ on the harmonics and the coefficient functions.

For $l=k=0$, the quantities multiplying $B$ and $C$ vanish, and
Eq.~(\ref{generalPertMetric}) reduces to our previous expression,
  Eq.~(\ref{ZeroEllMetric}).  If $l=1$, and thus $k^2=3$, the quantity
  multiplying $C$ vanishes as a result of the identity
\begin{equation}
     \partial_a \partial_b Y_{1 m}(\Omega)=-\bar{g}_{ab} Y_{1m}(\Omega)
\end{equation}
and there are only three metric
coefficients, $A$, $B$, and $\Psi$.  
 
Expanding to second order in these perturbations gives the quadratic 
Lagrangian
\begin{eqnarray}
     L^{(2)}_E&=&-\frac{3}{\kappa}\rho^3\dot{\Psi}^2
      -\frac{\rho}{\kappa}(k^2-3)\Psi^2
      +\frac{1}{2}\rho^3\dot{\Phi}^2
    +\frac{1}{2}\left(k^2\rho+\rho^3 U''\right)\Phi^2
     -3\rho^3\dot{\phi}\dot{\Psi}\Phi\,\cr
 &&+\frac{\rho^3}{3\kappa}\left(\frac{k^2-3}{k^2}\right)\dot{C}^2
  -\frac{\rho}{9\kappa}(k^2-3)C^2-\frac{2\rho}{3\kappa}(k^2-3)\Psi C \,\cr
  &&+\left[-\rho^3\dot{\phi}\dot{\Phi}+\rho^3 U'\Phi 
  +\frac{6}{\kappa}\dot{\rho}\rho^2\dot{\Psi}
  +\frac{6}{\kappa}(1-k^2/3)\rho\Psi+ \frac{2}{\kappa}(1-k^2/3)\rho C\right]A
 -\frac{3}{\kappa}\rho Q A^2 \,\cr
&&+\frac{2\dot{\rho} k^2}{\kappa} AB
   -\left[\frac{2\rho}{\kappa}\dot{\Psi}
   +\frac{2\rho}{3\kappa}\left(\frac{k^2-3}{k^2}\right)\dot{C}
   +\rho\dot{\phi}\Phi\right]k^2 B -\frac{k^2}{\kappa \rho}B^2 \, .
\label{GeneralL2}
\end{eqnarray}
This Lagrangian is invariant under the two-parameter coordinate transformation
\begin{eqnarray}
  \xi&\rightarrow& \xi+\alpha(\xi)Y(\Omega)   \, , \\
     z_a &\rightarrow& z_a + \beta(\xi)\partial_a Y(\Omega) \, ,
\end{eqnarray}
under which the perturbations transform as   
\begin{equation}
\delta_G \Phi = \dot{\phi}\alpha,\quad 
   \delta_G \Psi=\frac{\dot{\rho}}{\rho}\alpha-\frac{k^2}{3}\beta, \quad
   \delta_G A = \dot{\alpha}, \quad 
   \delta_G B=\alpha+\rho^2 \dot{\beta},\quad \delta_G C = k^2\beta \, .
\end{equation}

With two independent gauge parameters $\alpha$ and $\beta$ we expect to 
have two constraints.  Indeed, we see that no derivatives of $A$ or $B$
appear in $L^{(2)}$.  Differentiating with respect to these quantities
leads to the constraints
\begin{equation}
  0 ={\cal C}^{(1)}_A \equiv \frac{\kappa \rho^2}{6}
  \left(\dot{\phi}\dot{\Phi}-U'\Phi\right)
  -\left[\rho \dot{\rho}\dot{\Psi}+\left(1
  -\frac{k^2}{3}\right)\left(\Psi+\frac{C}{3}\right)
   +\frac{\dot{\rho}k^2}{3\rho}B\right]+ Q A 
\end{equation}
and 
\begin{equation}
 0 = {\cal C}^{(1)}_B \equiv \dot{\Psi}+\frac{k^2-3}{3k^2}\dot{C}
  +\frac{\kappa\dot{\phi}}{2}\Phi-\frac{\dot{\rho}}{\rho}A+\frac{1}{\rho^2}B=0 \, .
\end{equation}

Both of these constraints are gauge-invariant,
leading us to again seek a manifestly gauge-independent form of the 
Lagrangian.  Using the constraints to eliminate $A$ and $B$ and proceeding as 
before, we obtain 
\begin{equation}
  L^{(2)}_E=\frac{\rho^3(1-k^2/3)}{2(Q-\dot\rho^2k^2/3)} \dot{\chi}^2
   +\frac{\rho^3(1-k^2/3)}{2(Q-\dot\rho^2k^2/3)}f(\rho,\phi)\chi^2 \, ,
\label{L2forAllEll}
\end{equation}
where now $\chi$ is the only possible gauge-invariant generalization of
Eq.~(\ref{ZeroEllChi}),
\begin{equation}
   \chi\equiv \dot\rho\Phi- \rho\dot\phi\Psi
    -\frac{\rho\dot{\phi}}{3}C  \, ,
\end{equation}
and 
\begin{eqnarray}
   f &=&U''+ {k^2 \over \rho^2}  
   +\frac{\kappa}{3\dot\rho(Q-\dot\rho^2 k^2/3)}\left\{\rho\dot{\phi}U'
    + \rho^2 \dot\rho U'^2   
  -k^2\left[2 \rho \dot{\rho}^2 \dot{\phi}U'
    -3\dot{\rho}\dot{\phi}^2Q
   +\dot{\rho}\dot{\phi}^2\right]\right\}  \cr &&\qquad
   +\frac{2\kappa \dot\phi^2}{3} - \frac{\kappa\rho \dot\phi U'}{\dot\rho}
       -\frac{4\kappa U}{3}  -{\ddot \rho \over \dot \rho}
  \left({\dot Q - 2\dot\rho\ddot\rho k^2/3 \over Q - \dot\rho^2 k^2/3}\right) \, .
\end{eqnarray}
If $l =k=0$, Eq.~(\ref{L2forAllEll}) reduces to the
result of the previous subsection.  For $l=1$ and $k^2=3$, 
$L^{(2)}_E$ vanishes identically.  Finally, note that 
the coefficient of $\dot\chi^2$ is positive for all $l \ge 2$.

\subsection{Perturbations about a homogeneous solution}

The analysis of the previous two subsections is modified somewhat if
the unperturbed solution is homogeneous, with $\phi$ constant and
equal to an extremum of $U(\phi)$ and the metric being the usual round
metric on the four-sphere.  This could be the false vacuum solution
given by Eqs.~(\ref{HubbleDef}) and (\ref{SphereRho}) or the analogous
true vacuum solution, in either case with $U''>0$.  Alternatively, it
could be the Hawking-Moss solution~\cite{Hawking:1981fz} with
$\phi=\phi_{\rm HM}$ at a local maximum of $U$.

If we examine the quadratic Lagrangians in Eqs.~(\ref{zeroEllL2}) and
(\ref{GeneralL2}), we see that the terms coupling $\Phi$ with a metric
perturbation are all absent if $\dot\phi$ and $U'$ both vanish.
Furthermore, $\Phi$ drops out of the constraints, and $Q$ is
identically equal to unity.  Finally, $\chi$ is simply $\Phi$.
Redoing the analysis, we find that the metric terms are removed by the
constraints, and we obtain simply
\begin{equation}
    L^{(2)}_E = \frac12 \rho^3 \left[ \dot\Phi^2 + \left( U''
       +{k^2\over \rho^2}\right)\Phi^2 \right]  \, .
\end{equation}
(Note that the $l=1$ modes do not drop out, in contrast with
the perturbations about inhomogeneous bounces.)
The eigenmodes of this Lagrangian are just the five-dimensional 
spherical harmonics.  If
\begin{equation}
     \frac{U''}{H_{\rm top}^2} < -N(N+3)
\end{equation}
with $N=0,1,\dots$, then there are $N+1$ negative eigenvalues.
If these are numbered $n=0,1,\dots$, eigenvalue $n$ has a 
degeneracy $(2n+3)(n+2)(n+1)/6$.

\section{Negative modes with gravity}
\label{negmode-sec}

In this section we investigate the $\chi$-field 
modes with negative eigenvalues that arise from the
quadratic Euclidean actions that were obtained in
Sec.~\ref{perturb-sec}.  These actions, given in Eqs.~(\ref{g-inv-L2})
and (\ref{L2forAllEll}), are comprised of a ``kinetic energy'' term
quadratic in $\dot \chi$ and a ``potential energy'' term quadratic in
$\chi$.  For $l=0$ the kinetic energy can be either
positive or negative, depending on the sign of $Q$, while the sign
of the potential energy is that of $f/Q$.  For $l \ge 2$ the kinetic 
energy is always positive and the sign of the potential energy is
that of $f$.  (As already noted, there are no gauge-invariant modes 
for $l=1$.)

We can distinguish two classes of negative modes:

a) Slowly varying or ``standard'' negative modes: These are associated
with a positive kinetic energy and a negative potential energy. They
are the analogues of the negative mode about the flat-space bounce.  
We will restrict our investigation to the $l=0$ case; although we do 
not expect negative modes with higher $l$ (there are none in flat
spacetime), we have not been able to prove that these are impossible.

b) Rapidly oscillating modes: These arise if $l=0$ and $Q<0$, giving a negative
kinetic energy.  Low-amplitude, short-wavelength oscillations of the
form $\chi \sim\sin (\omega\xi)/\omega$ in the region of negative $Q$ 
can yield modes with negative eigenvalues whose magnitudes grow without
bound as $\omega$ tends toward infinity; this is a manifestation of the
conformal mode problem of Euclidean gravity~\cite{Gibbons:1978ac}.

We will consider these two cases separately.

\subsection{Slowly varying negative modes}

In flat spacetime these are associated with the variation of the
bubble radius $R$.  Indeed, in the thin-wall approximation, they are
signaled by the fact that $d^2S_E/dR^2$ is negative at $R=\bar R$, as
noted in Eq.~(\ref{TWAsecondDeriv}).  As has already been noted,
when the thin-wall approximation is applied to the case with
gravitational effects included, one finds that for a bubble with a
wall at $\rho=\bar\rho$ the second derivative $d^2 S_E/d\bar\rho^2$ is
negative for a type A de Sitter bounce, but positive for a type B
bounce.  This suggests that there is a slowly-varying negative mode in
the former case, but not in the latter.

To study this issue further, and going beyond the thin-wall limit, we
took as an example a theory with a scalar field potential,
\begin{equation}
    U(\phi) = (\phi-3)^2\phi^2 + 0.5\phi^2 + 1.5  \, ,
\label{firstPotential}
\end{equation}
whose false and true vacua are at $\phi= 2.81$ and $\phi = 0$,
respectively.  We studied the behavior of the bounce solutions as the
strength of gravity was progressively increased.  We worked with a
dimensionless $\phi$ and $\kappa$; the translation to physical
quantities is obtained by noting that for this potential the scalar
field mass scale $\mu$, defined in Eq.~(\ref{mudef}), is related to
the Planck mass by
\begin{equation}
    {\mu \over M_{\rm Pl}} = 0.24 \sqrt{\kappa} \, .
\end{equation}

\begin{figure}[t]
\centering
\begin{tabular}{cc}
\includegraphics[height=1.9in]{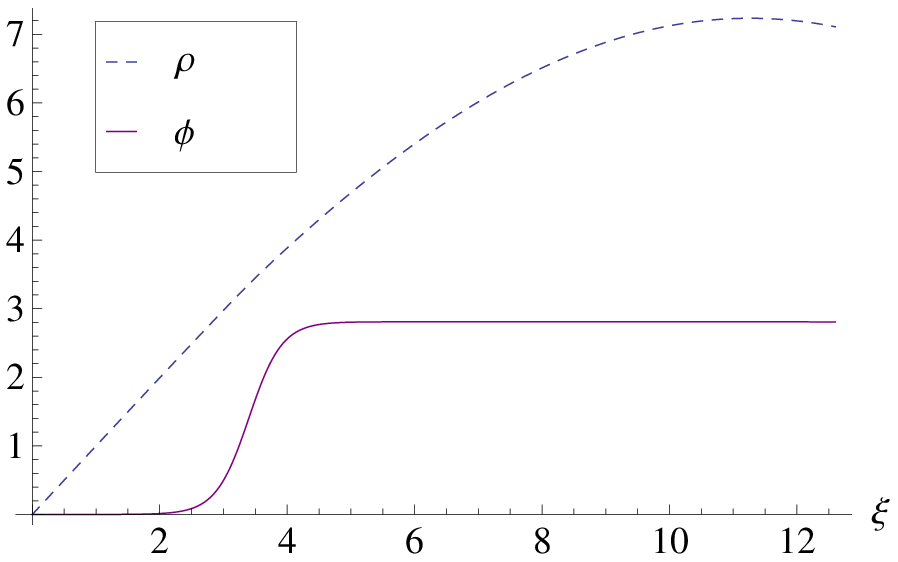}
\hskip .25in
\includegraphics[height=1.9in]{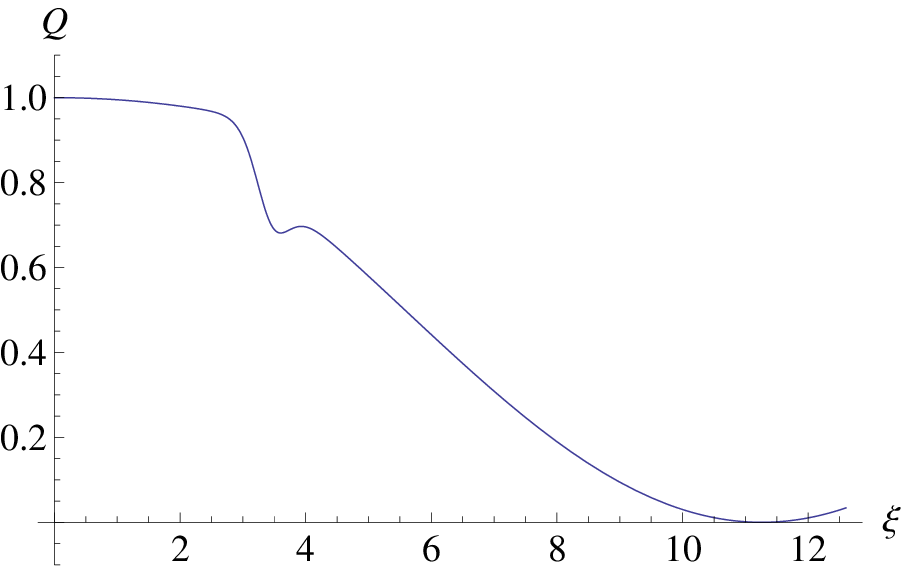} \\
   (a) \hskip 3.0in (b) \\
\includegraphics[height=1.9in]{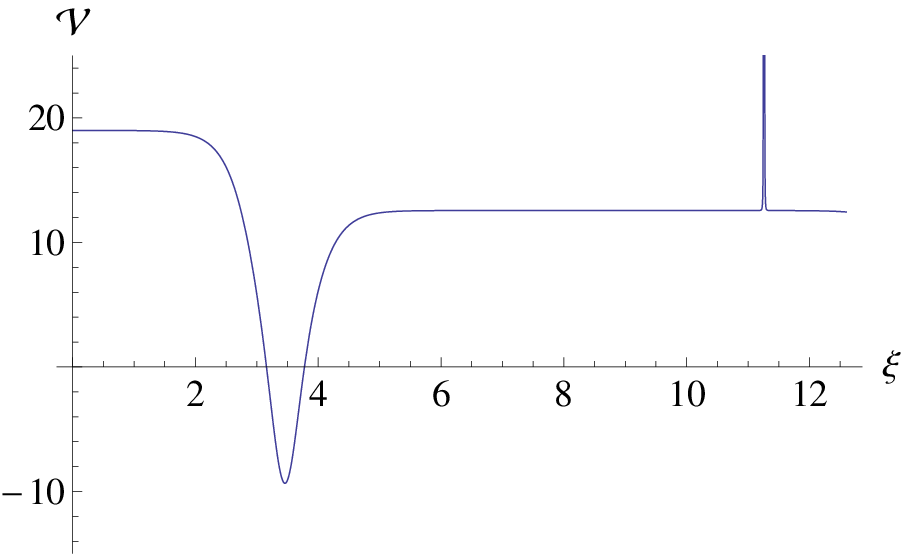}
\hskip .25in
\includegraphics[height=1.9in]{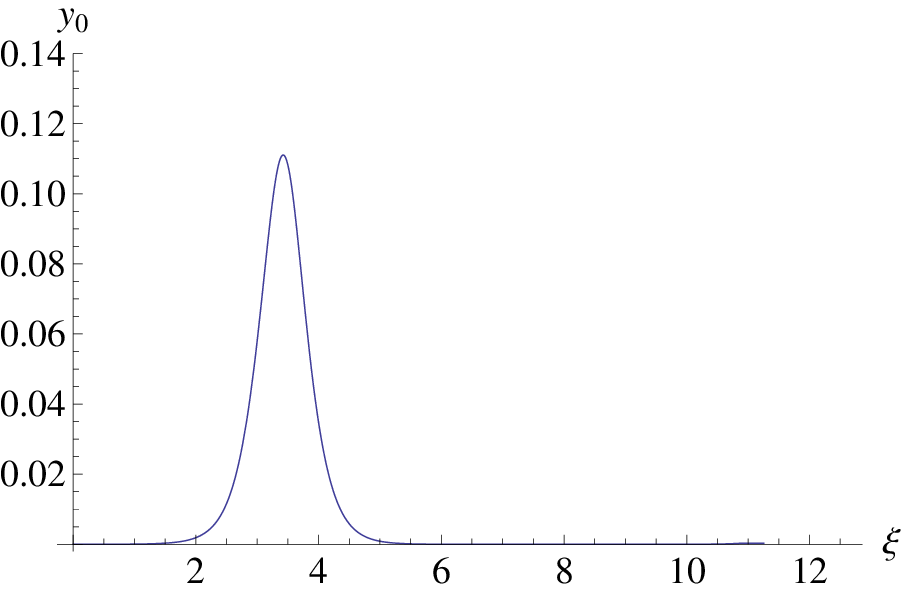} \\
   (c)
 \hskip 3.0in (d) \\
\end{tabular}
\caption{Results for the potential given in
  Eq.~(\ref{firstPotential}), with $\kappa=0.01$.  The bounce solution
  is shown in (a), with $\rho$ given by the dashed blue line and
  $\phi$ by the solid red line.  Panel (b) shows $Q$.  The minimum at
  the location of the maximum of $\rho$ is negative, although this is
  not apparent in the figure due to its exponentially small absolute
  value.  Panel (c) shows the function ${\cal V}$ given in
  Eq.~(\ref{calVdef}).  The lowest eigenmode of the Lagrangian of
  Eq.~(\ref{rewriteL}) is shown in (d); its eigenvalue is $-(0.40\mu)^2$.}
\label{kappa=01fig}
\end{figure}

\begin{figure}
\centering
\begin{tabular}{cc}
\includegraphics[height=1.9in]{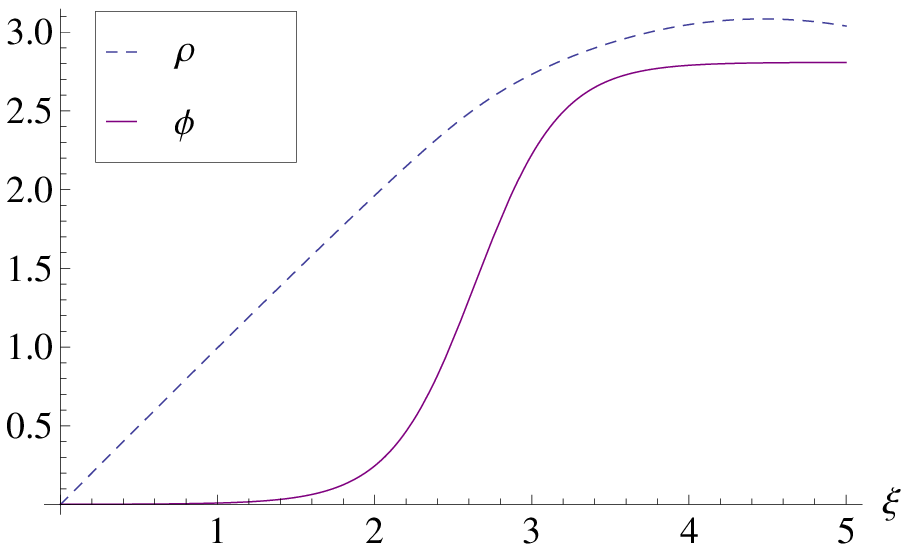}
\hskip .25in
\includegraphics[height=1.9in]{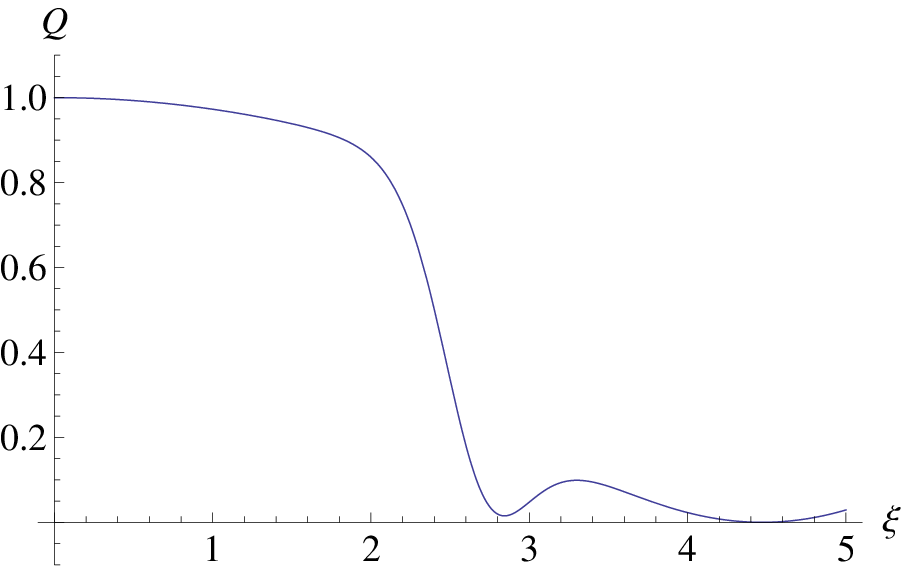} \\
   (a) \hskip 3.0in (b) \\
\includegraphics[height=1.9in]{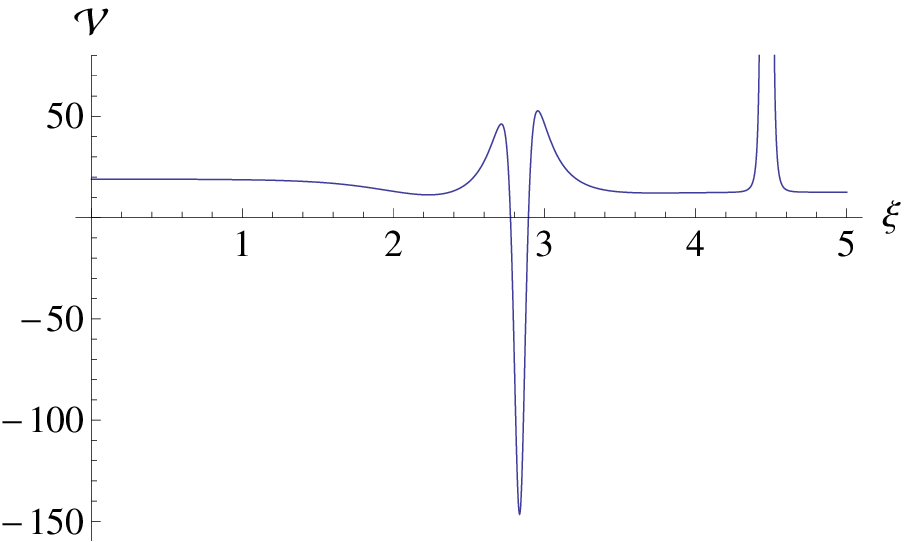}
\hskip .25in
\includegraphics[height=1.9in]{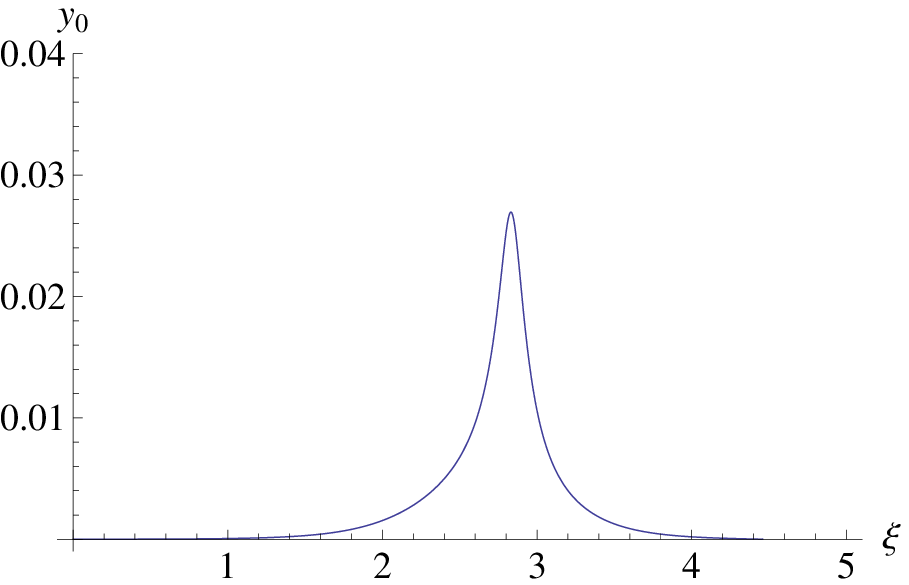} \\
   (c) \hskip 3.0in (d) \\
\end{tabular}
\caption{The same as in Fig.~\ref{kappa=01fig}, but for $\kappa=0.055$.  
At the first minimum $Q=0.015$, while at the second it has an exponentially 
small negative value.  The eigenvalue of the negative mode is $-(0.21 \mu)^2$.}
\label{kappa=055fig}
\end{figure}

To make the comparison with the flat-space case clearer we define
\begin{equation}
    y = \chi/\sqrt{Q}
\end{equation}
and rewrite Eq.~(\ref{g-inv-L2}) as
\begin{equation}
   L^{(2)}_E = {\rho^3\over 2}\left( \dot y^2 + {\cal V} y^2 \right) \, ,
\label{rewriteL}
\end{equation}
with 
\begin{equation}
    {\cal V} = U'' +{\kappa \rho^2 {U'}^2 \over 3  Q}
    +{\kappa \rho \dot\phi U' \over 3 \dot\rho Q} 
      -{3\dot\rho \dot Q\over 2\rho Q} - {\ddot Q \over 2Q}
      + {3 \dot Q^2 \over 4 Q^2}
      +{3\ddot\rho \over \rho} + {\dddot \rho \over  \dot \rho}
       -{\ddot\rho\dot Q\over \dot\rho Q}      \, .
\label{calVdef}
\end{equation}

\begin{figure}[t]
\centering
\begin{tabular}{cc}
\includegraphics[height=1.85in]{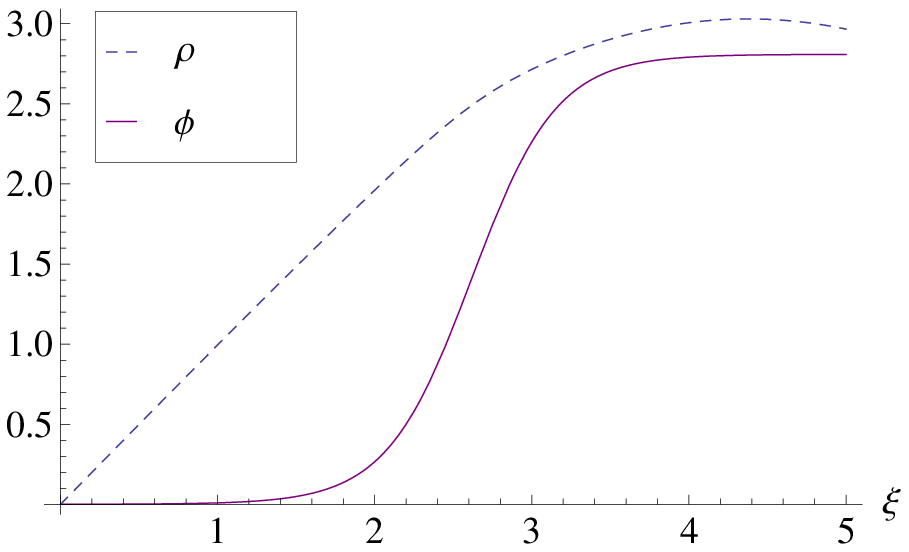}
\hskip .25in
\includegraphics[height=1.85in]{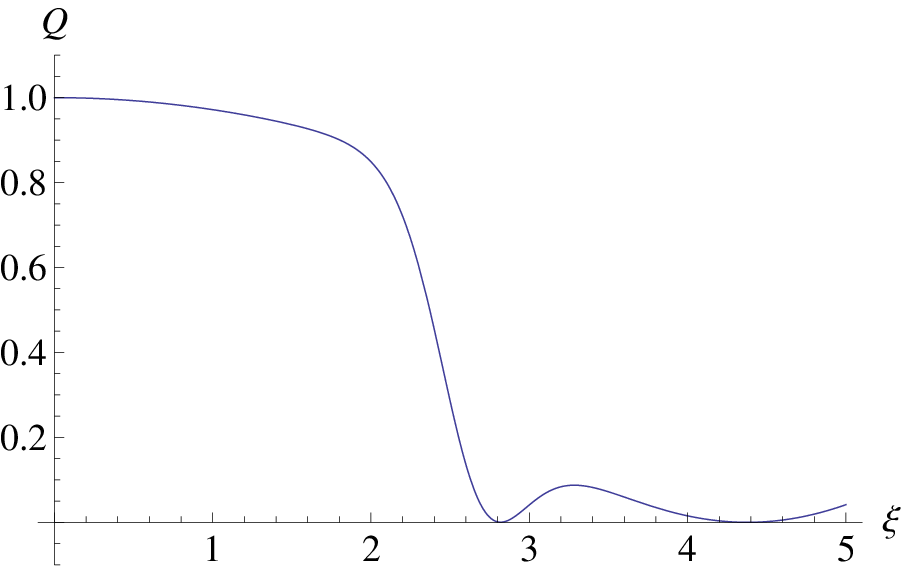} \\
   (a) \hskip 3.0in (b) \\
\includegraphics[height=1.85in]{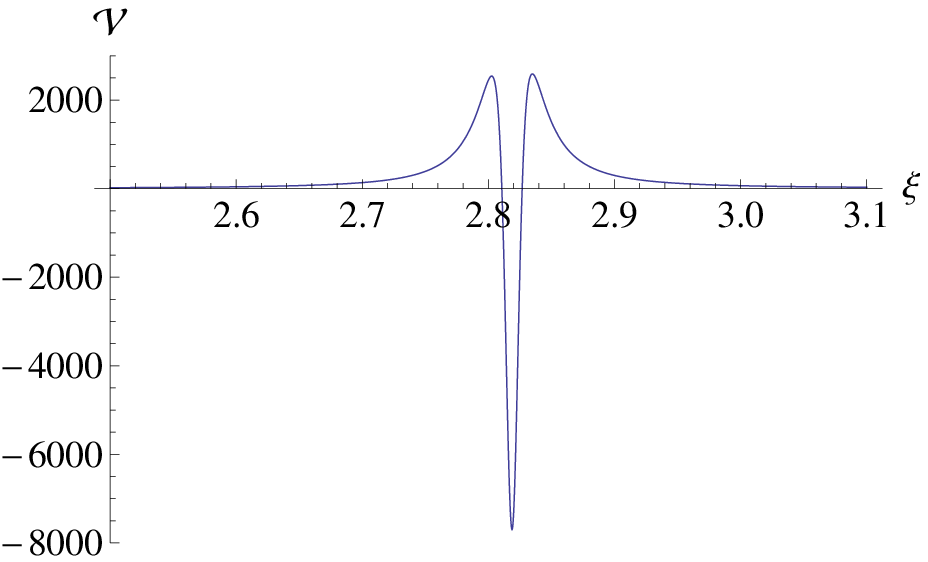}
\hskip .25in
\includegraphics[height=1.85in]{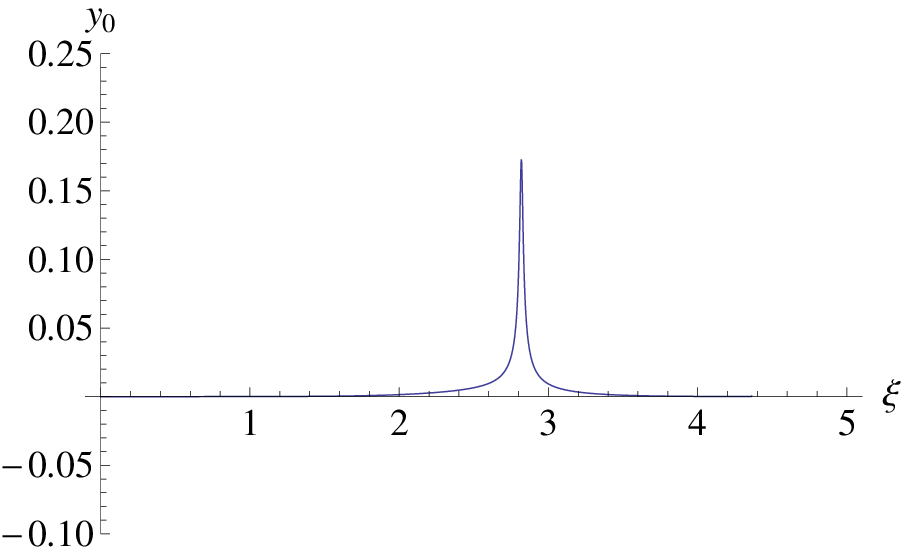} \\
   (c) \hskip 3.0in (d) 
% \\ \includegraphics[height=2.0in]{704e.eps} \\  (e)
\end{tabular}
\caption{Results for $\kappa=0.057$. The plots are analogous to those
  in Fig.~\ref{kappa=055fig}, except that in (c) the plot covers only
  a narrow range of $\xi$ containing the dip in ${\cal V}$.  Outside
  this range ${\cal V}$ shows little change from
  Fig.~\ref{kappa=055fig}. Again $Q$ has two minima, one positive and one 
negative; at the former $Q=0.00027$.  The lowest eigenmode, shown in 
(d), has a positive eigenvalue $(0.23 \mu)^2$. }
\label{kappa=057fig}
\end{figure}

\begin{figure}[t]
\centering
\begin{tabular}{cc}
\includegraphics[height=1.9in]{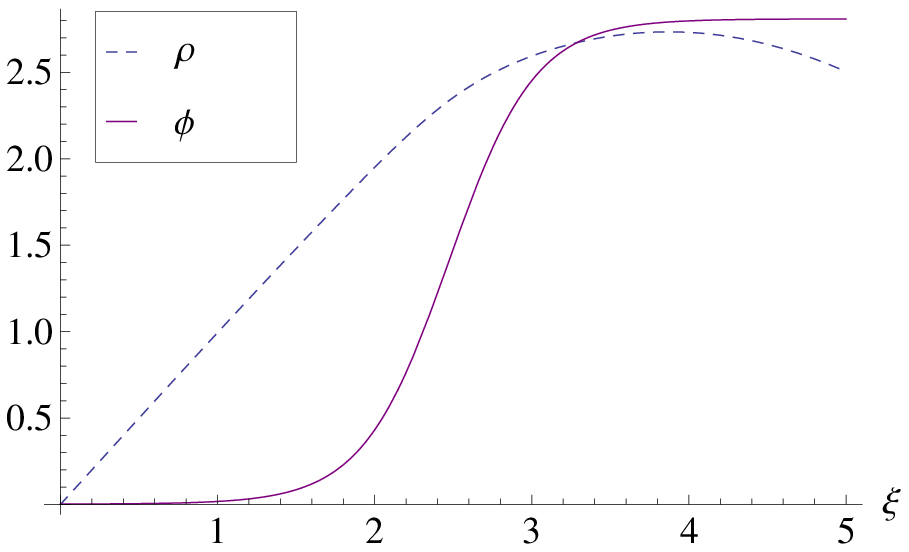}
\hskip .25in
\includegraphics[height=1.9in]{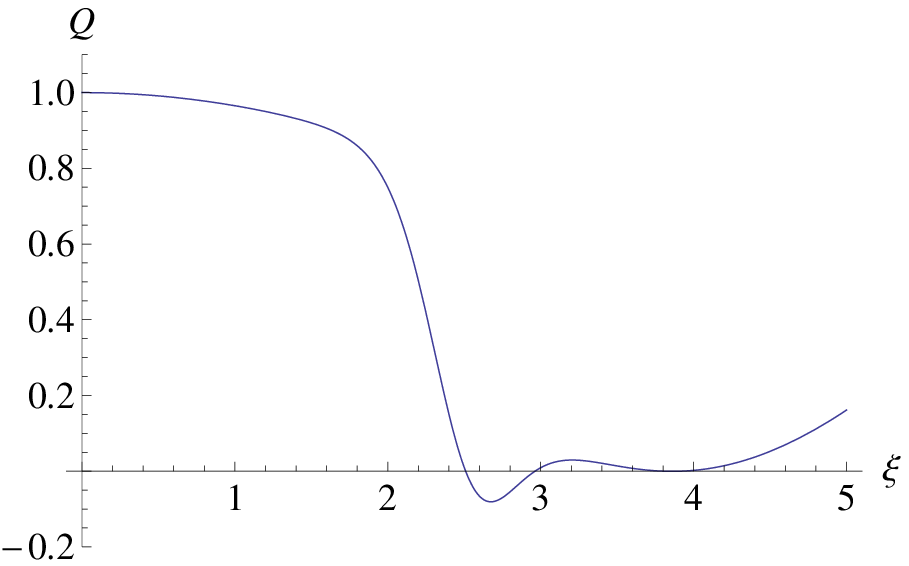} \\
   (a) \hskip 3.0in (b) 
\end{tabular}
\caption{Results for $\kappa=0.07$. Panels (a) and (b) are as in 
Fig.~\ref{kappa=01fig}.  Note that $Q$ now has two negative regions.}
\label{kappa=07fig}
\end{figure}

\begin{figure}
\centering
\begin{tabular}{cc}
\includegraphics[height=1.85in]{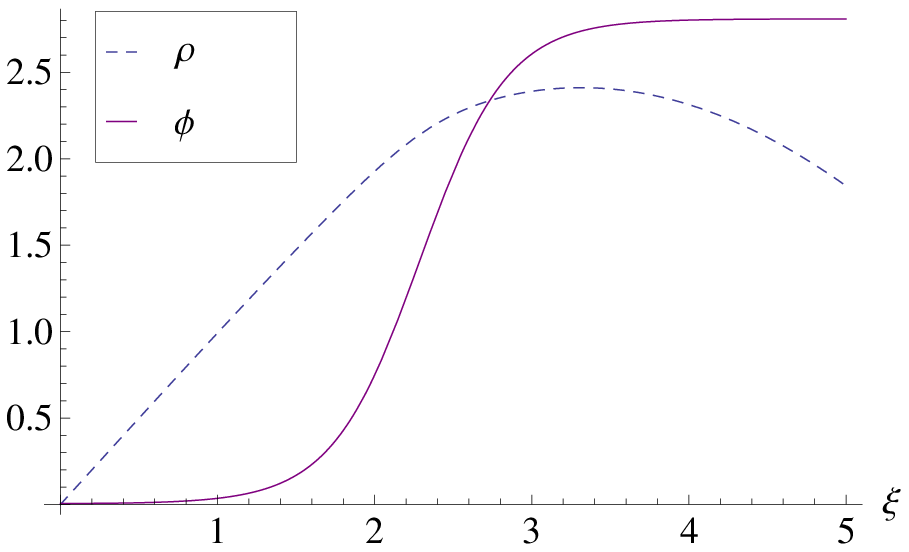}
\hskip .25in
\includegraphics[height=1.85in]{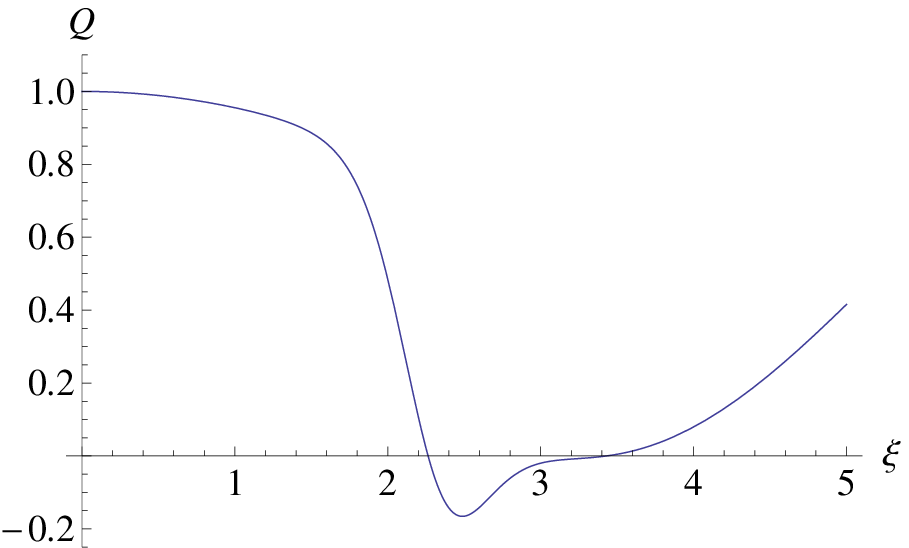} \\
   (a) \hskip 3.0in (b) 
\end{tabular}
\caption{Same as for Fig.~\ref{kappa=07fig}, but for $\kappa=0.09$.
  The regions of negative $Q$ have now merged to form a single
  connected region.}
\label{kappa=09fig}
\end{figure}

Figures~\ref{kappa=01fig} - \ref{kappa=09fig} show the evolution of
the bounce and the lowest eigenmode as the gravitational coupling
$\kappa$ is increased.  In Fig.~\ref{kappa=01fig} we show the bounce
fields $\phi(\xi)$ and $\rho(\xi)$ for the weak gravity case
$\kappa=0.01$ (i.e., $\mu = 0.024 M_{\rm Pl}$).  The bubble wall is
located well before the maximum of $\rho$, so this is clearly a type A
bounce, and might almost be classified as a small-bubble bounce.  The
dominant term in ${\cal V}$ is $U''$, as in flat space.  The negative
mode is concentrated on the wall region of the bounce.  Its eigenvalue
is $\lambda_0 = -(0.40 \mu)^2$, which can be compared with the
flat-space ($\kappa=0$) value $-(0.39 \mu)^2$.  There is a second
minimum of $Q$ at the value of $\xi$ where $\rho$ reaches its maximum.
From Eq.~(\ref{Qdef}) we see that $Q$ is negative there, although the
factor of $\dot\phi^2$ makes its absolute value exponentially small.
We will return to this minimum and the narrow region of negative $Q$
surrounding it in the next subsection.

Increasing the strength of gravity to $\kappa = 0.055$ (i.e., $\mu =
0.056 M_{\rm Pl}$) yields the results shown in
Fig.~\ref{kappa=055fig}.  The maximum of $\rho$ is outside the bubble
wall (although just barely so), and so this is a large type A bounce.
$Q$ remains positive throughout the wall region, although it reaches a
minimum value of 0.015 near the center of the bubble wall.
Gravitational effects have significantly modified ${\cal V}$, whose
negative region is much narrower.  
The negative mode has become somewhat narrower, but as
before it is centered about the minimum of $Q$ in the wall.
The eigenvalue has increased to $-(0.21 \mu)^2$.

The effect of a further increase in $\kappa$, to 0.057 (and, by a
numerical coincidence, $\mu = 0.057 M_{\rm Pl}$), is shown in
Fig.~\ref{kappa=057fig}.  The profiles of $\rho$ and $\phi$ show
little change, and the bounce remains type A.  The minimum value of
$Q$ in the wall remains positive, at 0.00027.  The dip in ${\cal V}$
is now much deeper and much narrower.  Most importantly, the lowest
eigenvalue is now positive, at $(0.23\mu)^2$, even though this is
still  a type A bounce.

With even a slight further increase of $\kappa$, the minimum value of
$Q$ on the bubble wall becomes negative.  This is illustrated in
Fig.~\ref{kappa=07fig} for the case $\kappa=0.07$ ($\mu= 0.063 M_{\rm Pl}$).
Because the change of
variables from $\chi$ to $y$ can no longer be carried out, $\cal V$
ceases to be a useful quantity, and there is no analogue of the lowest
eigenmodes of the previous examples.  If there is a slowly varying negative
mode, it is not simply related to the ones found in type A solutions with 
weaker gravity and smaller bounces.

With $\kappa$ increased to 0.09
($\mu= 0.072 M_{\rm Pl}$), we have a bounce, shown in Fig.~\ref{kappa=09fig},
that can be viewed as being on the borderline between type
A to type B.  We see that the regions of negative $Q$ at the wall and at
$\rho_{\rm max}$ have merged to form a single negative-$Q$ region.

\begin{figure}
\centering
\begin{tabular}{cc}
\includegraphics[height=1.9in]{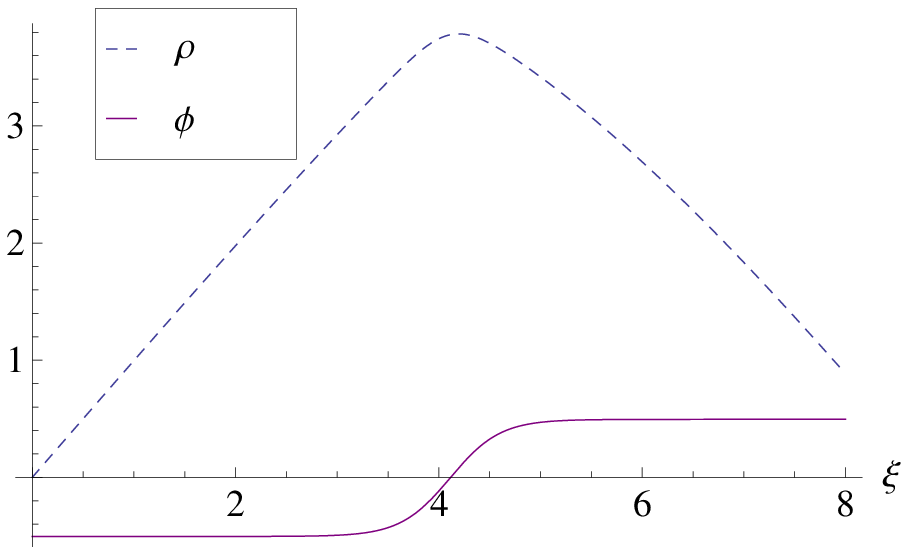}
\hskip .25in
\includegraphics[height=1.9in]{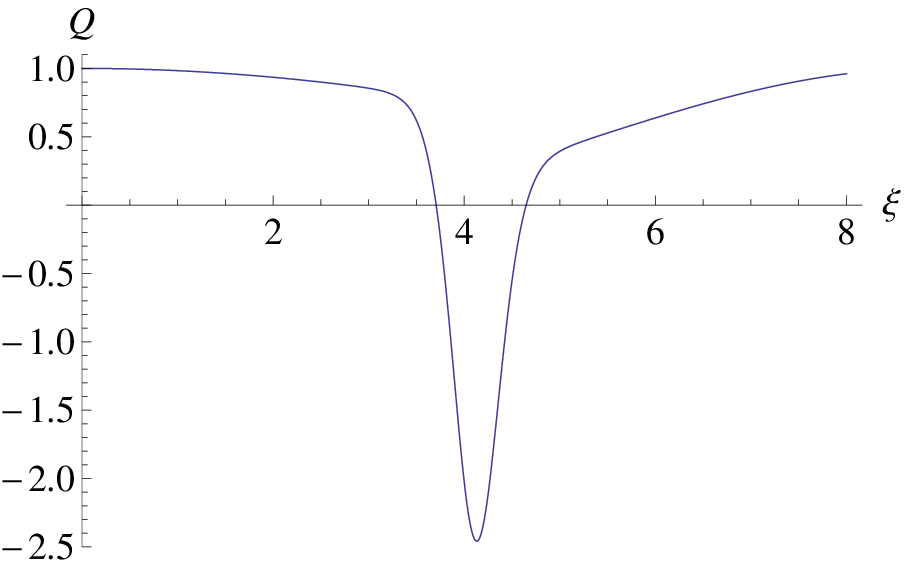} \\
   (a) \hskip 3.0in (b) \\
\includegraphics[height=1.9in]{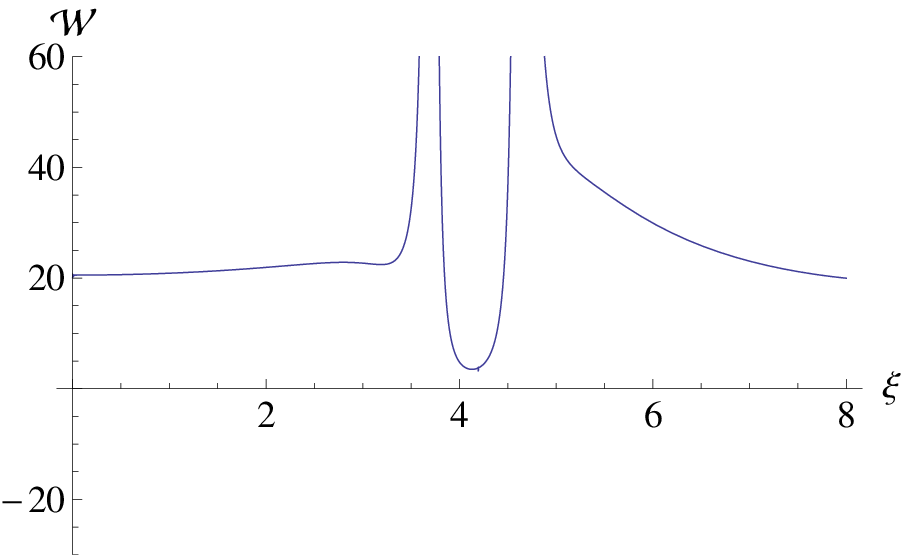} \\
     (c) 
\end{tabular}
\caption{A type B bounce, with $\kappa=1$ and the potential given by
  Eq.~(\ref{secondPotential}).  Panels (a) and (b) are as in
  Fig.~\ref{kappa=07fig}, while panel (c) shows ${\cal W}=f/Q$.  }
\label{typeBfig}
\end{figure}

Finally, in Fig.~\ref{typeBfig} we illustrate a clear-cut type B bounce.
Instead of Eq.~(\ref{firstPotential}), the potential is now given by
\begin{equation}
    U = 10 (\phi^2 - 0.25)^2 + 0.1(\phi+1)  \, .
\label{secondPotential}
\end{equation}
The plots are for $\kappa=1$, corresponding to $\mu/M_{\rm Pl} =
0.17$.  The potential ${\cal W} = f/Q$ is everywhere positive, so
there is no possibility of a standard slowly varying negative mode.

\subsection{Rapidly oscillating modes}

These arise whenever $Q$, defined by Eq.~(\ref{Qdef}), is negative; 
i.e., whenever
\begin{equation}
    {\kappa \rho^2 \dot\phi^2 \over 6 \dot\rho^2} > 1  \, .
\label{negQcondition}
\end{equation}
This can happen in two different situations.  For some bounces, $Q$
becomes negative in a portion of the bubble wall, leading to what we
will call ``wall modes''; we saw an example of this in
Fig.~\ref{kappa=07fig}.  For all bounces arising from de Sitter space,
$Q$ becomes negative in a region about
the maximum of $\rho$; the resulting negative modes may be called
``$\rho_{\rm max}$ modes''.  Of course, the distinction between these
classes only applies for type A bounces; for type B bounces, where
$\rho_{\rm max}$ is reached in the bubble wall, the distinction
disappears.  Indeed, Fig.~\ref{kappa=09fig} illustrates how the two
classes merge as type A crosses over to type B.

We start by considering a small bubble bounce, with $\mu \ll M_{\rm
  Pl}$ and the bubble radius $\bar \rho \ll H^{-1}$.  In the wall
region $\dot\phi^2 \sim \mu^4$, while $\dot\rho \approx 1$, so
\begin{equation}
    {\kappa \rho^2 \dot\phi^2 \over 6 \dot\rho^2} \sim 
       { \mu^4 \bar\rho^2 \over M_{\rm Pl}^2 }  \, .
\end{equation}
For a typical small bubble bounce this is much less than unity, and so
there are no wall modes.

There will, however, be $\rho_{\rm max}$ modes.  These will be 
centered about $\xi = (\pi/2H) - \Delta \equiv \xi_0$, where $\rho$ reaches
its maximum. From Eq.~(\ref{rhoInFV}) we find that in this region
$\rho\approx H^{-1}$ and 
\begin{equation} 
    \dot\rho \approx \sin[H(\xi_0-\xi)] \approx H(\xi_0-\xi)  \, ,
\end{equation}
while $\dot\phi$ is exponentially small, with $|\dot \phi| 
\sim \mu^2 e^{-\mu\pi/2H}$.  It follows that $Q$ is only negative in 
a region with an exponentially small width,
\begin{equation}
    \Delta \xi \approx {\mu^2 \over H^2 M_{\rm Pl}} 
      \,  e^{-\mu \pi/2H}  \, ,
\end{equation}
that is 
much less than a Planck length.  As a result, the
eigenvalues of the negative modes will all be super-Planckian.

We turn now to the case of a large type A bounce.  In the thin-wall 
approximation this has $\rho \approx H_t^{-1} \sin (H_t\xi)$ in 
the true vacuum region.  In order to have a region of negative $Q$, we
need that at the wall 
\begin{equation}
    {\kappa \rho^2 \dot\phi^2 \over 6 \dot\rho^2} \sim {\mu^4 \over H_t^2
      M_{\rm Pl}^2} \, \tan^2(H_t \bar\xi) \sim {\mu^4 \over U_{\rm
        tv}} \, \tan^2(H_t \bar\xi) 
\end{equation}
be greater than unity.  As we have already seen in
Fig.~\ref{kappa=07fig}, it is easy to construct examples where this is
true without invoking any Planckian mass
scales~\cite{Lavrelashvili:1985vn}.

These type A bounces will also have $\rho_{\rm max}$ modes, with
support in a region of width
\begin{equation}   
   \Delta \xi \approx {\mu^2 M_{\rm Pl}\over U_{\rm tv}}
    \,  e^{-\mu (\xi_0 - \xi)} \, .  
\end{equation}  
Even for a large type A bounce, such as those illustrated in
Figs.~\ref{kappa=055fig}-\ref{kappa=07fig}, where the wall is only a few
e-foldings away from $\rho_{\rm max}$, this width remains small,
although not necessarily sub-Planckian, almost until the point where the type A
bounce goes over to type B.

For type B bounces, where $\rho$ reaches its maximum inside the bubble
wall, the distinction between the two types of rapidly oscillating
negative modes disappears; a typical example is illustrated in
Fig.~\ref{typeBfig}.

Table~\ref{negmode-table} summarizes the results of this section.

\begin{table}
\begin{tabular}{|c|c|c|c|}
\hline
   & ~ Slowly varying ~  & ~ Wall oscillating ~  &~ $\rho_{\rm max}$ oscillating ~ \\
\hline
~Small type A~& Yes & No & Planckian  \\
~Large type A~ & Usually& Possible & Yes  \\
Type B & No & --- & Yes   \\
\hline
\end{tabular}  
\caption{Summary of the results of Sec.~\ref{negmode-sec}, indicating the types
of negative modes that occur with the various classes of bounces.  Note that
for type B bounces there is no distinction between the two types of 
oscillating negative modes.}
\label{negmode-table}
\end{table}

\section{Negative modes about multibounce configurations}
\label{multibounceSec}

In discussions of tunneling in flat spacetime, one usually focuses on the 
single-bounce solution and its associated determinant factor and normal 
modes.  In the usual dilute gas approximation, the deviation of the determinant
from its value in the pure false vacuum can be approximated as being local to
the bounce, so that for multibounce quasi-stationary points the determinant 
term, including its single factor of $i$, is simply repeated for each 
additional bounce.  Summing over all numbers of bounces leads to an exponential, 
with the factor of $i$ promoted to the exponent.

The situation is more subtle with gravitational effects included.  For
large type A bounces (and of course for all type B bounces), there may
not even be room on the Euclidean sphere to have several
well-separated bounces.  On the other hand, if the scalar field mass
scale $\mu$ is far below the Planck mass and the bounce radius is much 
less than $H^{-1}$, there is no problem at all 
with configurations containing large numbers of component bounces.  However, 
one wonders what becomes of the $\rho_{\rm max}$ negative modes, which are 
not localized about the bounce, when there are multiple bounces.

To start, consider an O(4)-symmetric solution with two bounces
centered at antipodal points\footnote{This is the simplest example of
  an oscillating bounce solution~\cite{Hackworth:2004xb}.}, which we
may take to be $\xi=0$ and $\xi=\xi_{\rm max}$.  There is potentially
a negative-$Q$ region about the ``equator'' at $\xi=\xi_{\rm max}/2$.
By symmetry, $\dot\phi=0$ here, so both terms in 
\begin{equation}
     Q = \dot\rho^2 - {\kappa \rho^2 \dot\phi^2 \over 6}
\end{equation}
vanish at the equator.  If the bounces have radii much less than the
horizon length, $\dot\phi$ will be exponentially small, $Q$ will be
positive, and the infinite set of oscillating negative modes will be
absent.  On the other hand, with larger bounces one can easily find
parameters that would make the first term smaller than the second in
the region close to the equator, giving a negative $Q$ and a family of
oscillating negative modes.

\begin{figure}
\centering
\includegraphics[height=4.0in]{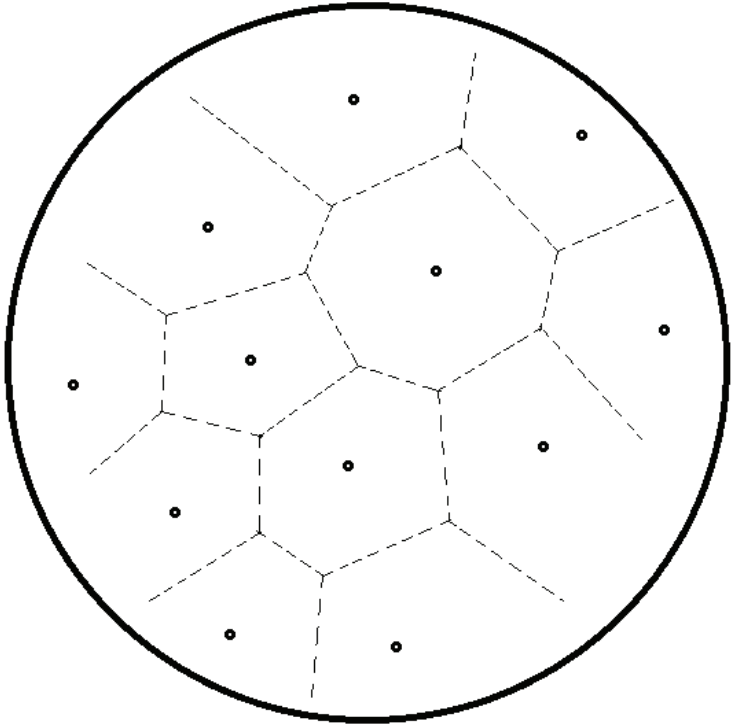}
\caption{Schematic illustration of a many-bounce configuration.}
\label{manybubblesfig}
\end{figure}

Having seen that two-bounce solutions may or may not have negative-$Q$
regions, let us turn to the more generic multibounce case, with many
component small bounces.  A schematic view of such a solution is shown
in Fig.~\ref{manybubblesfig}.  Here the small circles represent the
individual bounces, with $\phi$ close to its true vacuum value near
their centers.  As one moves away from one of these circles $\phi$
rapidly approaches $\phi_{\rm fv}$.  Thus, in most of the space $|\phi
- \phi_{\rm fv}|$ is exponentially small.  The magnitude of this
exponential tail decreases as the distance to the nearest bounce
increases.  Roughly speaking, this tail decreases in magnitude until
one reaches the dotted lines in the figure, which schematically
represent the boundaries that separate the ``domains'' of the
individual bounces.

Although we have no overall O(4) symmetry to guide us, we can look to
our analysis of the single bounce case for guidance.  We begin by
remarking that the fields in the interior of each domain are, to a
good approximation, the same as those in a region of comparable size
about the bounce in a one-bounce solution.  Since the rapidly
oscillating negative modes in the one-bounce solution have support far
from this region, we should not expect to find them in the domain
interiors.

Hence, if rapidly oscillating negative modes exist, they should be
located near the domain boundaries. However, even that seems quite
unlikely.  With many component bounces, the typical domain size is
much smaller than the curvature radius of the underlying four-sphere.
Thus, viewed on the scale of a few domains, the solution is barely
distinguishable from the corresponding flat-space multibounce
configuration, which we know has no oscillating negative modes.  To
phrase this differently, in the one-bounce solution the negative-$Q$
region is a thin three-sphere shell with a
curvature radius equal to that of the four-sphere background.  In the
multibounce configuration the domain boundaries approximate thin
three-sphere shells with curvature radii much smaller than that of the
underlying background.

We conclude that while rapidly oscillating negative modes exist for 
the one-bounce solution, and possibly even for quasi-stationary points
with a handful of bounces, they will be absent for configurations 
with many bounces.  We need to ask ourselves which of these cases 
is the one of relevance for us.

In the discussion of tunneling in flat spacetime in
Sec.~\ref{flatTunnelSec} we argued that the path integral was
dominated by configurations with $n$ bounces, with $n$ given by
Eq.~(\ref{dominantN}).  With a total four-volume of $\Omega T$, we had
$n \sim (\Omega T) \Gamma$.  Of course, the four-volume was understood
to be taken to infinity at the end of the calculation, so the
conclusion was that the path integral was dominated by configurations
in which the bounces were infinite in number but with a finite density
of the order of $\Gamma$.

If we ignore for the moment any complications from rapidly varying
negative modes (which will, in any case, be absent if $n$ is large),
then analogous reasoning leads to the conclusion that the dominant
configurations for de Sitter tunneling should be ones with
\begin{equation}
    n \approx \left({8\pi^2 \over 3}\, H^{-4}\right) \left(K e^{-B}\right) 
\end{equation}
component bounces.  The first factor, ${8\pi^2 \over 3}\, H^{-4}$, is the four-volume
of the Euclidean de Sitter space.  The second factor is (apart from a 
factor of 2) the usual
expression for the nucleation rate per unit volume, $\Gamma$.
Dimensional analysis suggests that
\begin{equation}
     r \equiv {\Gamma \over H^4} 
     \sim \left({M_{\rm Pl}\over \mu} \right)^4 \, e^{-B} \, .
\end{equation}
Depending on the parameters of the scalar field theory, this 
ratio could easily be much less than or much greater than unity;
given the exponential dependence on the bounce action, obtaining 
a value very close to unity, while possible, is unlikely.

If $r \gg 1$ the path integral is overwhelmingly dominated by
configurations with many bounces.  These configurations have only the
usual slowly varying negative modes, one for each bounce.  The
anomalous rapidly varying negative modes are absent.  On the other
hand, if $r \ll 1$ the contributions from multibounce configurations
are negligible, and the single-bounce solution, with its infinite set
of negative modes, makes the important contribution to the path
integral.\footnote{The situation here is somewhat akin to that with
  the ``interior'' B-to-C bounces discussed in Sec. 3.  In that case,
  the finite Euclidean volume of the A-to-B bounce led to two limiting
  regimes, one with no B-to-C bounces and one with many.  Here the
  finite volume of the Euclidean spacetime implies limiting cases
  with one and with many CDL bounces.}

\section{Discussion and summary}
\label{conclusionSec}

In examining various CDL bounces, we have found patterns of negative
modes that are rather different from those encountered in the absence
of gravity.  It seems natural to ask why matters aren't the same as in
flat spacetime.  However, one might instead ask why they should be the
same.  The issue is not so much the formal difference between the two
cases, with constraints and gauge issues involved in one case and 
not the other, but rather the fact that different physical processes come into
play when gravitational effects are included.

This can be seen already in the case of small bounces, with all mass
scales far below the Planck mass and the bounce radius much less than
the horizon distance.  In the absence of gravity we have a standard
first-order transition.  Once the transition begins (e.g., by cooling
below a critical temperature) small bubbles start to nucleate in the
initial false vacuum.  These bubbles then expand, collide, and
coalesce to form a homogeneous true vacuum.  Varying the magnitude of 
$\Gamma$ changes the time scale for completion of the transition, but
makes no qualitative change in the process.

With gravity brought into play, and a de Sitter false vacuum, matters
are more complicated.  If $r=\Gamma/H^4 \gg 1$, the time scale for
bubble nucleation, expansion, and coalescence is short compared to the
Hubble time and the transition to a homogeneous new phase is completed
very much as in flat spacetime.  On the other hand, if $r \ll 1$, the
cosmic expansion outpaces the bubble nucleation and the transition to
the true vacuum is never truly completed~\cite{Guth:1981uk}.  More
precisely, if $r$ is less than a critical value $r_c$ (which can be
shown to lie in the range $1.1 \times 10^{-6} < r_c < 0.24$), then the
new phase never percolates~\cite{Guth:1982pn}.  Instead, false vacuum
regions always remain, continuing to nucleate new bubbles of the true
vacuum, and eternal inflation ensues.

The underlying explanation for the difference between the two cases is
the manner in which the bubble grows in de Sitter space.  After
nucleating with a small initial radius, the bubble expands at a speed
that quickly approaches the speed of light, with the bubble wall
approximately tracing out a light cone.  On time scales much less than
$H^{-1}$, there is very little difference from the flat-space case; if
$\Gamma$ is large enough for the transition to have been completed by
this time, nothing unusual is found.  However, once $t\sim H^{-1}$ and
the bubble radius approaches the horizon length, the cosmic expansion
dominates and the bubble radius, as measured in comoving coordinates,
becomes essentially constant.  Hence, bubbles that are initially
separated by a distance greater than $2H^{-1}$ never meet.  Bubbles
that nucleate later have smaller asymptotic comoving sizes; they
nucleate and expand in the spaces left between the older bubbles, but
never quite fill out these false vacuum regions.

The bounce solution itself shows no distinction between the regimes of
large and small $r$.  In both cases the CDL bounce approaches the
flat-space solution as $\kappa \rightarrow 0$.  However, differences
appear when we look at the pre-exponential factor.  The two regimes
correspond to the two cases found in Sec.~\ref{multibounceSec}.  If $r
\gg 1$ the path integral is dominated by configurations with many
bounces.  These have the usual expansion/contraction negative modes,
one for each component bounce, but no indication at all of any
anomalous rapidly varying negative modes.  If instead $r \ll 1$, only
the single-bounce configuration, with its rapidly oscillating negative
modes near $\rho_{\rm max}$, is relevant.

While these $\rho_{\rm max}$ modes do not appear in some Hamiltonian
choices of gauge and, because their eigenvalues are super-Planckian,
may well disappear with a proper treatment of quantum gravity, there
is another class of modes that also distinguishes between the two regimes.
As we saw in Eq.~(\ref{L2forAllEll}), the $l=1$ modes make no contribution to the
quadratic fluctuation Lagrangian about a single bounce.  The
zero-eigenvalue $l=1$ modes simply correspond to translation of the
bounce, and must be replaced by collective coordinates even in flat
spacetime.  However, the flat-space $l=1$ modes with nonzero
eigenvalue are true fluctuation modes, and it is puzzling that they
do not have direct analogues about the single de Sitter CDL bounce. 
With the multibounce configurations matters are different.  The low-lying
fluctuations in the neighborhood of each component bounce, including
the ones that locally look like $l=1$ modes, closely approximate
their flat-space counterparts.  

Thus, for relatively rapid nucleation, with $r\gg 1$, the physical
processes proceed very much as in flat spacetime, and the calculation
of $\Gamma$, including the pre-exponential factor, smoothly
transitions between the gravitational and the non-gravitational cases.  If
instead $r \ll 1$, the physical process in the presence of gravity is
qualitatively different from when $\kappa=0$, and this difference is
reflected in the calculation of the pre-exponential factor.  Note that
$\rho_{\rm max}$, where the potential pathologies in the calculation
arise, corresponds to the location of the horizon in the initial and
final states.  This fits nicely with the fact that it is the existence
of a horizon that is responsible for the failure of the transition to
truly complete.
  
It is instructive to recall that the basis of the path integral
calculation was the interpretation of the imaginary part of the energy
of the false vacuum in terms of the lifetime of an unstable false
vacuum state.  On relatively small scales, with or without gravity, we
have a region of space that is initially in a homogeneous false vacuum
state and then tunnels to a state in which a true vacuum bubble is
embedded in a surrounding false vacuum region.  On a larger scale,
matters are more subtle.  Without gravity, or with gravity and $r \gg
1$, the continued bubble nucleation leads to a complete transition to
the true vacuum; the false vacuum unambiguously decays.  On the other
hand, if $r \ll 1$, an eternal inflation scenario results and the
false vacuum never fully disappears; in a sense, its lifetime is
infinite.

Let us turn now to the case of large bounces, but, as always, with all
mass scales well below the Planck scale.  Several physical processes
are possible.  First, we can have nucleation of a single bubble that,
even initially, occupies a large fraction of a horizon volume.  Given
the size of the bubble, the possibility of multibounce configurations
is severely constrained.  Second, there can be bounces that occupy a
full horizon volume and that are naturally interpreted as mediating
transitions that occur over the entire horizon volume.  These include
type B bounces, with a horizon volume of one vacuum tunneling into a
horizon volume of a different vacuum, and Hawking-Moss bounces, with
an entire horizon volume thermally fluctuating to the top of the
potential barrier.  We might also include here the oscillating bounce
solutions.  Although it has been argued that these are not relevant
for vacuum transitions because they have multiple negative modes even
when the gravitational background is held
fixed~\cite{Hackworth,Lavrelashvili:2006cv,Battarra:2012vu,Dunne:2006bt}, it may
be that this conclusion should be revisited.  Finally, there can be
``up-tunneling'', with a bubble of a higher (false) vacuum forming
within a region of lower (true) vacuum~\cite{Lee:1987qc}.

Issues of time-slicing arise when considering these bubbles.  In
discussions of vacuum tunneling one often speaks as if the bubble
nucleates suddenly and simultaneously over a hypersurface of constant
time, with the classical evolution and expansion of the bubble
beginning sharply at some $t=t_0$.  Even ignoring the neglect of the
quantum fuzziness, already in flat spacetime there is an ambiguity due
to the possibility of Lorentz boosting the nucleation hypersurface,
although this is known to not affect the large time evolution of 
the bubble~\cite{Garriga:2012qp,Frob:2014zka}.  

This issue becomes more acute with de Sitter bubble nucleation,
especially for the case of large bounces.  For example, if the de
Sitter spacetime is described in global, closed-universe coordinates,
then one can always find a de Sitter transformation that brings the
center of the bounce to a point on the $t=0$ ``waist'' of the de
Sitter hyperboloid and the nucleation hypersurface to this
hypersurface.  Indeed, this is the simplest choice for
continuing the Euclidean solution to a Lorentzian one.  From this
viewpoint, the horizon volume occupies half of the full $t=0$ de
Sitter space, and the nucleated bubble a large fraction of that half.
Alternatively, one could use flat-universe coordinates (which cover
half of the fully extended de Sitter spacetime).  With this choice a
spacelike hypersurface of fixed time is infinite in extent and the
initial horizon volume, and the nucleated bubble within it, appear as
just small regions within a larger universe.  This latter viewpoint
fits much better with the idea that the vacuum transition should not
be sensitive to the global structure of spacetime.

Even with a given choice of slicing, there are ambiguities in defining
particles, and thus the vacuum state, for the true vacuum, and perhaps
even more so for the false vacuum.  The analysis of
Ref.~\cite{Brown:2007sd} derived the CDL bounce by working within the
static coordinate patch of a horizon volume.  The vacuum naturally corresponding
to these coordinates is not the same as, e.g, the Bunch-Davies vacuum.  Does the
choice of one or the other affect the value of the imaginary part of
the vacuum energy?

Focusing now on the negative modes, we have both unanticipated modes
that are present, and an expected mode that is absent.  The former
include the formally infinite family of $\rho_{\rm max}$ oscillating
modes that, unlike those associated with small bounces, can have
sub-Planckian eigenvalues.  As with the corresponding small bounce
modes, the fact that $\rho_{\rm max}$ corresponds to the horizon fits
well with the fact that the existence of the horizon plays a
significant role in determining the nature of the transition governed
by these bounces.

Also unanticipated are the oscillating negative modes with support on
the bubble wall.  In our analysis these appear when $\kappa \rho^2
\dot\phi^2/6 > \dot\rho^2$; with the Hamiltonian gauge choice used in
Refs.~\cite{Khvedelidze:2000cp,Gratton:2000fj,Dunne:2006bt}, they
require the only slightly stronger condition that $\kappa \rho^2
\dot\phi^2/6 > 1$.  We find these modes puzzling.  The condition for
their existence does not entail any unusually large local energy
density.  Rather, the large circumference of the bounce, a nonlocal
property, seems to be crucial.

The expected mode that goes missing is the standard slowly varying
mode corresponding to radial expansion or contraction of the bounce.
This is absent from the thin-wall analysis for all type B bounces (but
see Ref.~\cite{Yang:2012cu}).  Going beyond the thin-wall
approximation, our numerical results showed the mode deviating more
and more from its expected form, eventually merging with the rapidly
oscillating negative modes, as the bounce moved from type A toward
type B.  

Is this a problem?  Is it an indication that type B bounces do not
correspond to vacuum transitions?  We think not.  Recall from the
discussion below Eq.~(\ref{TWArhoBar}) that the type B regime is the
one where $\epsilon$, the energy density difference of the two vacua,
is small or even zero.  Bubble nucleation certainly occurs for large
positive $\epsilon$, and we know that the thermal nature of de Sitter
spacetime allows the very same bounces to mediate tunneling upward;
i.e., with large negative $\epsilon$~\cite{Lee:1987qc}.  
Continuity arguments then
strongly suggest that tunneling with $\epsilon \approx 0$ should also
be possible.  However, this tunneling need not be a decay.  A single particle
in a double-well potential can tunnel back and forth between two 
degenerate minima.  The corresponding Euclidean solution has no negative
mode.  Perhaps the type B bounces should be understood as describing 
a finite system --- the fields within a horizon volume --- tunneling 
between two degenerate states.

Let us sum up briefly.  We have investigated the negative mode problem of 
de Sitter CDL bounces  in a manner somewhat complementary to previous 
studies, with the goal of gaining physical insight into the 
various anomalies and pathologies associated with these modes.

For the case of small bounces, we have found that the presence or
absence of potential oscillating negative modes, as well as of the
$l=1$ modes, depends on the value of $\Gamma/H^4$.  There are two regimes,
which correspond to the two possible outcomes of de Sitter vacuum decay
--- either a rapidly completed transition or eternal inflation.  Only
the former goes over smoothly to the flat spacetime case, showing that
the true weak gravity limit of bubble nucleation requires not only
that the mass scales be well below the Planck mass and the initial
bubble radius much less than the horizon distance, but also that the
characteristic time scale for nucleation be small compared to
$H^{-1}$.

For large bounces, we found further evidence, confirming the
indications from thin-wall arguments, that type B bounces do not have
the usual expansion/contraction negative mode.  We argued that,
nevertheless, these bounces correspond to vacuum transitions.  On the
other hand, the underlying physical origin of the oscillating negative
wall modes that arise when the bounce is sufficiently large remain
somewhat obscure.  Elucidating this issue is among the problems
that remain to be resolved by future investigations.

\begin{acknowledgments}

We are grateful for the hospitality of the Korea Institute for
Advanced Study, where part of this research was performed.  This work
was supported in part by U.S. Department of Energy grants
DE-FG02-92ER40699 and DE-SC0011941.

\end{acknowledgments}

\appendix

\section{Horizon volumes and the bounce solution}
\label{horizonApp}

In this appendix we show that the the slices of the Coleman-De Luccia
bounce that we identified in Fig.~\ref{CDL-bounce} as giving the
configurations at the beginning and end of tunneling are indeed
horizon volumes.

The O(4)-invariant Euclidean metric can be written as 
\begin{equation}
 ds^2 = d\xi^2 +\rho(\xi)^2 \left( d\chi^2 +\cos^2\chi \,d\Omega_2^2\right) \, .
\end{equation}
The three-dimensional slices in which we are interested are given by $\chi=0$,
and either $0  \le \xi <  \xi_0$ or $ \xi_0 < \xi < \xi_{\rm max}$, 
where $ \xi_0$ is the location of the maximum of $\rho$.  

This Euclidean metric can be continued to a Lorentzian one along the 
hypersurface $\chi=0$ by writing $\chi = it$ and going to real $t$.  This
gives the metric
\begin{eqnarray}
   ds^2 &=&  - \rho^2 \, dt^2 + d\xi^2 + \rho^2 \cosh^2t \,d\Omega_2^2  \cr
   &=& -\rho^2 \, dt^2 + {1 \over {\dot\rho}^2 }\, d\rho^2 
         + \rho^2 \cosh^2t \,d\Omega_2^2  \, .
\end{eqnarray}
We see that on the $t=0$ hypersurface $\rho$ coincides with the usual
choice for radial coordinate.  The vanishing of $g^{\rho\rho} = \dot\rho^2$
at $\xi = \xi_0$ indicates the presence of a horizon, as claimed, 
and verifies that the two three-dimensional slices each correspond to 
a horizon volume in an instantaneously static metric.

In fact, because the bounce is a solution of the
Euclidean equations, we can use Eq.~(\ref{rho-eq}) and write
\begin{eqnarray}
     g^{\rho\rho} &=& 1 +{\kappa \over 3} \rho^2 \left[ \frac12 \dot\phi^2 
         -U \right]   \cr
       &=& 1 - {\kappa {\cal M}(\rho) \over 4\pi \rho}  \, .
\end{eqnarray}
Here we have defined 
\begin{equation}
    {\cal M}(\rho) = {4\pi \over 3} \, \rho^3 \left[ \frac12 \dot\phi^2
         -U \right]  \, .
\end{equation}
Differentiating it with respect to $\rho$, we obtain
\begin{eqnarray}
   {d{\cal M} \over d\rho} &=& 4\pi \left\{\rho^2 \left[ \frac12 \dot\phi^2 - U\right]
    +\frac13\rho^3 \left[\dot\phi\ddot\phi - {dU\over d\phi}\dot\phi\right]{1 \over \dot\rho}
      \right\} \cr
   &=&  -4\pi \rho^2 \left[ \frac12 \dot\phi^2 + U\right] \, ,
\end{eqnarray}
where the last line is obtained with the help of Eq.~(\ref{phi-eq}).  This, together
with the fact that ${\cal M}$ vanishes at the 
zeros of $\rho$, allows us to write
\begin{equation} 
     {\cal M}(\rho) = 4\pi \int_0^\rho d\rho \, \rho^2 \left[\frac12 g^{\rho\rho} 
    \left({d\phi \over d\rho}\right)^2 + U \right]  \, ,
\end{equation}
which is the familiar result for a static spherically symmetric field configuration.
A similar result follows for the slice with $\xi_0 < \xi <\xi_{\rm max}$.

\section{Perturbation Lagrangian}
\label{mode-app}

In this appendix we outline the steps leading to Eq.~(\ref{g-inv-L2})
for the quadratic Lagrangian governing O(4)-invariant fluctuations about the 
bounce.  

Simple substitution of the perturbations defined in
Eqs.~(\ref{zeroEllL2}) and (\ref{PhiPert}) into the unperturbed
Lagrangian gives the expression in Eq.~(\ref{ZeroEllMetric}), which can 
be written as 
\begin{equation}
    L^{(2)}_E =
    -\frac{3}{\kappa}\rho^3 \dot{\Psi}^2+
    \frac{3}{\kappa}\rho\Psi^2+\frac{1}{2}\rho^3\dot{\Phi}^2
    +\frac{1}{2}\rho^3 U''\Phi^2-3\rho^3 \dot{\phi}\dot{\Psi}\Phi
       + FA   -\frac{3}{\kappa}\rho  Q A^2  \, ,
\label{FullAction}
\end{equation}
where 
\begin{equation}
    F= -\rho^3\dot{\phi}\dot{\Phi}+\rho^3 U' \Phi
      +\frac{6}{\kappa}\dot{\rho}\rho^2\dot{\Psi}
    +\frac{6}{\kappa}\rho \Psi \, .
\end{equation}
Varying $L^{(2)}_E$ with respect to $A$ leads to the constraint equation
\begin{equation}
    F = {6\over \kappa} \, \rho Q A  \, .
\end{equation}
Using this to eliminate $A$, we can rewrite the Lagrangian in a form, 
\begin{equation}
    L_E^{(2)} =  -\frac{3}{\kappa}\rho^3 \dot{\Psi}^2+
    \frac{3}{\kappa}\rho\Psi^2+\frac{1}{2}\rho^3\dot{\Phi}^2
    +\frac{1}{2}\rho^3 U''\Phi^2-3\rho^3 \dot{\phi}\dot{\Psi}\Phi +
    {\kappa \over 12}\, {F^2 \over \rho Q}  \, ,
\end{equation}
that involves only $\Phi$ and $\Psi$.  

Now let us define a field
\begin{equation}
    Y = \Phi - {\rho \dot\phi \over \dot\rho} \, \Psi
\end{equation}
that is invariant under the infinitesimal gauge transformation of
Eq.~(\ref{O4gaugeTrans}).

Substituting this into our previous expression gives
\begin{eqnarray}
     L_E^{(2)} &=& {\rho^3 \over 2Q} \left[\dot\rho^2  \dot Y^2 
         + \left(U'' Q +{\kappa\over 6}\rho^2 U'^2 \right)Y^2
         - {\kappa\over 3}\rho^2 \dot\phi \,U' \dot Y Y\right]  +\cdots  \cr
     &=& {\rho^3\over 2Q} \left[\dot\rho^2  \dot Y^2
   + \left(U'' + {\kappa \rho^2 U'^2\over 3Q} 
     + {\kappa\rho \dot\phi U'\over 3\dot\rho Q}
    \right)\dot\rho^2 Y^2 \right]
  +\cdots \, .
\label{actionWithDots}
\end{eqnarray}
In the first line the ellipsis denotes terms containing $\Psi$ or $\dot\Psi$,
while in the second it also includes total derivative terms arising from 
the integration by parts.

Now recall that the action is gauge-invariant.  In particular, one
can verify that under an infinitesimal gauge transformation the
expression in Eq.~(\ref{FullAction}) only changes by a total derivative,
\begin{equation}
    \delta_G L_E^{(2)} =  {d \over d\xi}  \left[\left( 
  \rho^3 \ddot\phi \Phi +{6\over \kappa}\rho\Psi\right)\alpha\right] \, ,
\end{equation}
that makes no contribution to the total action.  When the action is
rewritten in terms of $Y$ and $\Psi$, the terms shown explicitly in
Eq.~(\ref{actionWithDots}) are manifestly gauge-invariant.  By
contrast, the omitted terms involving $\Psi$ or $\dot\Psi$ cannot
possibly be gauge-invariant, and so must in fact vanish.

Finally, noting that $Y$ is potentially singular when $\dot\rho=0$, let
us define
\begin{equation}
    \chi = \dot\rho Y = \dot\rho \Phi - \rho \dot\phi \Psi \, .
\end{equation}
Substituting this into Eq.~(\ref{actionWithDots}), performing an integration 
by parts to eliminate the terms proportional to $\chi \dot\chi$, and dropping 
total derivatives and terms involving $\Psi$ then leads to the expression
given in Eqs.~(\ref{g-inv-L2}) and (\ref{fForL2}).

\end{document}